%% file: main.tex
\def\IEEETRANS{}
\def\SHORTFORMDATASETS{}
\def\TWOAPRSCENARIOS{}

\ifdefined\LNCSCONF
\documentclass[runningheads]{llncs}

\else
\ifdefined\IEEETRANS
\documentclass[lettersize,journal,compsoc]{IEEEtran}
\else
\documentclass[sigconf,silent]{acmart}
\setcopyright{acmcopyright}
\copyrightyear{2021}
\acmYear{2021}
\acmDOI{}

\fi
\fi

\input{macros}
\def\BibTeX{{\rm B\kern-.05em{\sc i\kern-.025em b}\kern-.08em
    T\kern-.1667em\lower.7ex\hbox{E}\kern-.125emX}}

\begin{document}
\title{\papertitle 
}
\ifdefined\IEEETRANS
\author{\IEEEauthorblockN{Pemma Reiter\IEEEauthorrefmark{1},
Hui Jun Tay\IEEEauthorrefmark{1},	
Westley Weimer\IEEEauthorrefmark{2},
Adam Doup\'{e}\IEEEauthorrefmark{1},
Ruoyu Wang\IEEEauthorrefmark{1},
Stephanie Forrest\IEEEauthorrefmark{1}}

\IEEEauthorblockA{\IEEEauthorrefmark{1}~Arizona State University,
\IEEEauthorrefmark{2}~University of Michigan}
}
\else
\ifdefined\LNCSCONF

\author{
{Pemma Reiter}\inst{1}
\and
{Hui Jun Tay}\inst{1}
\and
{Westley Weimer}\inst{2}
\and
{Adam Doup\'{e}}\inst{1}
\and
{Ruoyu Wang}\inst{1}
\and
{Stephanie Forrest}\inst{1}
}

	\institute{Arizona State University \\
		\email{\{pdreiter,htay2,doupe,fishw,steph\}@asu.edu}
		\and
		University of Michigan
		\email{weimerw@umich.edu}
	}


\else

\fi
\fi

\iffalse
\thispagestyle{empty}
\else
\thispagestyle{plain}
\pagestyle{plain}
\fi

\ifnum\ifdefined\LNCSCONF 1\else 0\fi\ifdefined\IEEETRANS 1\else 0\fi=1

\maketitle
\begin{abstract} 
	\input{abstract}

\end{abstract}

\else

\begin{abstract} 
	\input{abstract}
\end{abstract}
\maketitle

\fi


\input{introduction}

\input{outline}




\ifdefined\LNCSCONF
\bibliographystyle{splncs04}
\else
\ifdefined\IEEETRANS
\bibliographystyle{IEEEtran}
\else
\bibliographystyle{ACM-Reference-Format}
\fi
\fi

\input{references}

\end{document}

%% file: macros.tex
\usepackage{amsmath,amsfonts}
\usepackage{algorithmic}
\usepackage{graphicx}
\usepackage{verbatim}
\usepackage{cite}

\ifdefined\IEEETRANS
\usepackage{algorithm}
\usepackage{array}
\usepackage[caption=false,font=scriptsize,labelfont=sf,textfont=sf]{subfig}
\usepackage{textcomp}
\usepackage{stfloats}
\usepackage{url}
\else
\usepackage{caption}
\usepackage[lofdepth,lotdepth,caption=false]{subfig}
\usepackage{booktabs}
\usepackage{longtable}
\usepackage{algorithm2e}
\usepackage{algpseudocode}
\usepackage[export]{adjustbox}

\fi

\usepackage{textcomp}
\usepackage{xcolor}
\usepackage{hyperref}
\usepackage{framed}

\usepackage{enumitem}
\usepackage[normalem]{ulem}
\usepackage{xspace}

\DeclareMathAlphabet{\mathpzc}{OT1}{pzc}{m}{it}
\usepackage{xurl}

\usepackage{listings,multicol}
\usepackage{xcolor}
\definecolor{mygreen}{rgb}{0.6,0.8,0.6}
\definecolor{myred}{rgb}{0.8,0.6,0.6}

\definecolor{codegreen}{rgb}{0,0.6,0}
\definecolor{codegray}{rgb}{0.5,0.5,0.5}
\definecolor{codepurple}{rgb}{0.58,0,0.82}
\definecolor{backcolour}{rgb}{0.95,0.95,0.92}
\definecolor{backcolour2}{rgb}{0.92,0.92,0.95}

\definecolor{mywhite}{rgb}{1,1,1}

\newcommand{\sysacronym}{BinrePaiReD\xspace}
\newcommand{\papertitle}{Automatically Mitigating Vulnerabilities in Binary Programs via Partially Recompilable Decompilation}

\newcommand{\ourgit}{\small \url{git@github.com:pdreiter/FuncRepair.git}}

\newcommand{\PartialRecompilation}{Partial recompilation\xspace}

\newcommand{\binarymanipulation}{binary rewriting\xspace}

\newcommand{\AutomatedProgramRepair}{Automated Program Repair\xspace}
\newcommand{\automatedprogramrepair}{automated program repair\xspace}
\newcommand{\APR}{APR\xspace}

\newcommand{\PartiallyRecompilableDecompilation}{Partially Recompilable Decompilation\xspace}
\newcommand{\PRD}{PRD\xspace}

\newcommand{\SECTION}[1]{\section{#1}}
\newcommand{\SUBSECTION}[1]{\subsection{#1}}
\newcommand{\SUBSUBSECTION}[1]{\subsubsection{#1}}

\newcommand{\SUBSUBSUBSECTION}[1]{\paragraph{\small #1}}

\newcommand{\FRAMED}[1]{\vspace{-0.25cm}\begin{framed}\noindent{#1}\end{framed}\vspace{-0.5cm}}

\newcommand{\circled}[1]{{\scriptsize\textcircled{\textbf{\textit{\tiny #1}}}}}

\graphicspath{{figures/}}

\usepackage{array}
\usepackage{color, colortbl}
\usepackage{hhline}
\definecolor{Gray}{gray}{0.95}
\definecolor{DGray}{gray}{0.93}
\definecolor{LGray}{gray}{0.97}
\definecolor{Green}{rgb}{0.72265625,0.875,0.64453125}

\definecolor{CFLLred}{RGB}{250,220,217}
\definecolor{CFLred}{RGB}{250,187,180}
\definecolor{CFred}{RGB}{250,144,134}

\definecolor{CFLLorange}{RGB}{252,221,199}
\definecolor{CFLorange}{RGB}{252,188,151}
\definecolor{CForange}{RGB}{252,145,98}

\newcolumntype{C}{>{\centering\arraybackslash}m{1.65cm}}
\newcolumntype{X}{>{\centering\arraybackslash}m{1cm}}
\newcolumntype{L}{>{\raggedleft\arraybackslash}m{2cm}}
\newcolumntype{Y}{>{\raggedleft\arraybackslash}m{1.25cm}}
\newcolumntype{R}{>{\raggedright\arraybackslash}m{1.5cm}}
\newcolumntype{Z}{>{\raggedright\arraybackslash}m{1.25cm}}

\newcolumntype{O}[1]{>{\raggedright\arraybackslash}p{#1}}
\newcolumntype{P}[1]{>{\centering\arraybackslash}p{#1}}
\newcolumntype{Q}[1]{>{\raggedleft\arraybackslash}p{#1}}

\newcolumntype{G}[1]{>{\columncolor{DGray}\raggedleft\arraybackslash}p{#1}}
\newcolumntype{g}[1]{>{\columncolor{LGray}\raggedleft\arraybackslash}p{#1}}
\newcolumntype{H}[1]{>{\columncolor{DGray}\centering\arraybackslash}p{#1}}
\newcolumntype{h}[1]{>{\columncolor{LGray}\centering\arraybackslash}p{#1}}

\lstdefinestyle{taupeCstyle}{
	backgroundcolor=\color{backcolour},   
	commentstyle=\color{codegreen},
	keywordstyle=\color{magenta},
	numberstyle=\tiny\color{codegray},
	stringstyle=\color{codepurple},
	basicstyle=\ttfamily\scriptsize,
	breakatwhitespace=false,         
	breaklines=true,                 
	captionpos=t,                    
	numbers=left,                    
	numbersep=5pt,                  
	showspaces=false,                
	showstringspaces=false,
	showtabs=false,
	frame=single,  
	tabsize=2
}
\lstdefinestyle{grayCstyle}{
	backgroundcolor=\color{backcolour2},   
	commentstyle=\color{codegreen},
	keywordstyle=\color{magenta},
	numberstyle=\tiny\color{codegray},
	stringstyle=\color{codepurple},
	basicstyle=\ttfamily\scriptsize,
	breakatwhitespace=false,         
	breaklines=true,                 
	captionpos=t,                    
	numbers=left,                    
	numbersep=5pt,                  
	showspaces=false,                
	showstringspaces=false,
	showtabs=false,                  
	frame=single,  
	tabsize=2
}
\lstdefinestyle{whiteCstyle}{
	backgroundcolor=\color{mywhite},  
	commentstyle=\color{codegreen},
	keywordstyle=\color{magenta},
	numberstyle=\tiny\color{codegray},
	stringstyle=\color{codepurple},
	basicstyle=\ttfamily\scriptsize,
	breakatwhitespace=false,         
	breaklines=true,                 
	captionpos=b,                    
	keepspaces=true,                 
	numbers=left,                    
	numbersep=5pt,                  
	showspaces=false,                
	showstringspaces=false,
	showtabs=false,                  
	tabsize=2
}
\newif\ifthree

\threetrue

%% file: abstract.tex
\label{sec:abstract}
Vulnerabilities are challenging to locate and repair, especially when source code is unavailable and binary patching is required.  
Manual methods are time-consuming, require significant expertise, and do not scale to the rate at which new vulnerabilities are discovered.  
Automated methods are an attractive alternative, and we propose Partially Recompilable Decompilation (\PRD). 
\PRD lifts suspect binary functions to source, available for analysis, revision, or review, 
and creates a patched binary using source- and binary-level techniques.
Although decompilation and recompilation do not typically work on an entire binary, our approach succeeds because it is limited to a few functions, like those identified by our binary fault localization.

We evaluate these assumptions and find that, without any grammar or compilation restrictions, 70-89\% of individual functions are successfully decompiled and 
recompiled with sufficient type recovery. In comparison, only 1.7\% of the full {\small C}-binaries succeed.
When decompilation succeeds, \PRD produces test-equivalent binaries 92.9\% of the time. 

In addition, we evaluate \PRD  
in two contexts: a fully automated process incorporating source-level Automated Program Repair (APR) methods; human-edited source-level repairs.  
When evaluated on DARPA Cyber Grand Challenge (CGC) binaries, 
we find that \PRD-enabled \APR tools, operating only on binaries, performs as well as, and sometimes better than full-source tools, 
collectively mitigating 85 of the 148 scenarios, a success rate consistent with these same tools operating with access to the entire source code.
\PRD achieves similar success rates as the winning CGC entries, sometimes finding higher-quality mitigations than those produced by top CGC teams.  
For generality, our evaluation includes two independently developed APR tools and \texttt{\small C++}, Rode0day, and real-world binaries.

%% file: introduction.tex
\section{Introduction}
\label{sec:introduction}
\IEEEPARstart{F}ixing software bugs
is challenging when vendor support is unavailable, source code is not provided, or rebuilding the software from source is infeasible.
In these cases, bugs must be mitigated at the binary level, which is a tedious, complicated, and error-prone task that does not easily scale to the considerable number of bugs plaguing today's software.  
At the binary level, the most compelling bugs are security vulnerabilities, which are important to address quickly after disclosure to reduce attack and exploitation. \footnote{This work has been submitted to the IEEE for possible publication. Copyright may be transferred without notice, after which this version may no longer be accessible.}

One path for addressing this challenge is automated program repair (\APR). 
Despite active development on many different tools and techniques,
most \APR methods today operate only on source code~\cite{repair-living-review}.  
An appealing alternative could lift the entire binary to source, where it can be analyzed, modified, 
and then recompiled to provide a patched binary.  
Unfortunately, current decompilation tools suffer scalability issues~\cite{Liu_Wang_2020} or focus on readability rather than recompilability~\cite{Botacin_Galante_de_Geus_Gregio_2019,Schulte_Ruchti_Noonan_Ciarletta_Loginov_2018,yakdan2015}, often producing inaccurate or non-compilable results when applied to a whole binary~\cite{Liu_Wang_2020}. 

Instead, this paper presents a hybrid approach to automated binary repair. 
Our approach centers on the idea that for most (if not all) binary programs, partial analysis is sufficient for binary repair. 
These insights guided our approach: 
fault localization can identify a small set of functions relevant to the vulnerability; 
decompilers can lift a small set of functions to recompilable source code; 
binary-source interfaces and binary rewriting can transform them into test-equivalent binaries, even when tools fail for full binaries;
the set of decompiled binary functions provide sufficient context to enable source-level analyses and transformations, even when those methods only operate on source.

Practically, we provide a mechanism through which partial analysis content can be consolidated to achieve automated patching in an effective manner.
However, the underlying technique is laborious, as existing tools are non-existent and generating compatible binary content from source, altering compiled binaries with that content, and ensuring that the result remains executable are all difficult.
Our fully-functional prototype addresses these difficulties by automatically resolving the burden of patching source into binaries. 
Specifically, it analyzes multiple abstractions and generates binary-source interfaces so that decompiler output can be used.
To suit our method, only the offset and referenced types for the decompiled function need recovery, significantly reducing the requirement of complete and sound type inference on binary code, an open research problem.
We call our decompilation-based, binary-source patching method Partially Recompilable Decompilation (PRD) and its \emph{source-based \APR}-compatible extension to automated binary repair, \sysacronym.
Our evaluation focuses on our tool's applicability to automated binary repair, its generality to real-world vulnerabilities, other languages, and performance constraints, as well as its inherent assumptions. 
Previous evaluations of decompiler output have limited source code, by restricting grammar and types~\cite{Liu_Wang_2020}, we evaluate decompiled code's recompilablility and behavioral consistency without these restrictions. 
As automated binary repair is dependent upon fault localization, we gauge our technique's effectiveness identifying functions relevant to vulnerabilities.
Our evaluation includes \texttt{\small C} and \texttt{\small C++} x86 programs from multiple datasets,  
two independently developed and evaluated \APR tools (Prophet~\cite{long2016automatic} and GenProg~\cite{le_goues_genprog_2012}), and off-the-shelf tools (e.g., Hex-Rays and GCC).
Our datasets include the DARPA Cyber Grand Challenge (CGC)~\cite{Song_Alves-Foss_2015}, Rode0Day~\cite{rode0day}, and
\texttt{\small C} and \texttt{\small C++} programs from MITRE CVE. 
Amazingly, these \APR tools perform well on these datasets despite being employed on a small set of functions formed from decompiled source code.
Our implementation can extend to other architectures and stripped binaries, assuming decompiler support. 

\smallskip
\noindent
To summarize, the main contributions of this paper are:
\begin{itemize}[topsep=2pt]
\item \PRD: a fully functional prototype that enables binary repair using high-level source code, the first such to support source-level patching of a binary.
\item An empirical validation of our assumptions by evaluation of the individual techniques comprising our solution. 
Without any grammar or compilation restrictions, we find that 70-89\% of individual functions can successfully decompile and 
recompile with sufficient type recovery (while only 1.7\% of full binaries succeed). PRD produces test-equivalent binaries 92.9\% of the time.

\item An {end-to-end} evaluation of \sysacronym's ability to apply source-based \APR tools to mitigate vulnerabilities in CGC binaries.
We find that two APR tools, when used with \PRD, mitigate vulnerabilities with success rates comparable to, sometimes exceeding, these same tools operating on the full source.  They collectively mitigate 85 of 148 unique defects (20 that the winning cyber-reasoning system failed to patch) and provide code that humans can analyze and amend.

\item Case studies that demonstrate language (C and C++) and compiler (Clang and GNU) generality, as well as no significant impact to performance.
\end{itemize}
\noindent
To further reproducible science, our prototypes, datasets, and our experimental results are available at
{{\ourgit}}.\label{sec:pkg}

%% file: outline.tex


\input{background}

\input{motivate}

\input{technical}

\input{dataset}

\input{evaluation}

\input{discussion}

\input{related}

\input{conclusion}


%% file: background.tex
\SECTION{Background}
\label{sec:background}

We briefly describe the techniques \PRD uses for analyzing and manipulating binary content, and those \sysacronym employs to localize and repair faults.

\SUBSECTION{\textbf{Binary Decompilation and Rewriting.}}
\emph{A binary program}, or \emph{binary}, refers to a structured executable file composed of encoded binary instructions.
\emph{Disassembling} is the process of lifting binary instructions to assembly;
\emph{decompilation} is the general process of lifting lower-level abstractions (e.g., assembly) to high-level representations, (e.g., source code).
Since much information, like control flow structures, prototypes, variable names, and types, is lost during compiling, decompilers must infer this content~\cite{yakdan2015,gussoni2020,loginov2016}.
Such inference is unsound, often leading to unreadable output, incorrect results, or failures.
\emph{Binary rewriting} directly alters a compiled binary file, retaining its ability to execute~\cite{Larus_Schnarr_1995}.
\PRD uses a binary rewriting strategy that appends new content to an existing binary, overwriting with additional calling requirements to \textit{detour} execution to new binary content (see Section~\ref{sec:prd_tech}). 

\SUBSECTION{\textbf{Fault Localization.}}
\label{subsec:fl}
\emph{Fault localization} (FL) methods  pinpoint the likely locations of a
vulnerability or bug by analyzing a program, dynamically or statically. In Spectrum-Based Fault Localization (SBFL), program spectra (characteristics) are obtained through analysis
and used to implicate code regions. Spectra can include code
coverage, data- or
information-flow~\cite{masri2010fault,de2014data}, call
sequences~\cite{zhu2017fault}, or program counter
samples~\cite{Schulte_DiLorenzo_Weimer_Forrest}.  
SBFL produces suspiciousness scores and ranking (risk evaluation)~\cite{abreu2009practical} and does not use other content, like historical development information~\cite{hirsch2020fault}. 
To locate vulnerabilities, SBFL map lower-level code elements, like statements and variables, to their execution, then apply metrics.
Most methods require access to source, but we localize to function spectra using only the binary and available test contents, calculating 
suspiciousness scores using multiple SBFL metrics.
Specifically, we adapt a hybrid FL method, Rank Aggregation
Fault Localization (RAFL)~\cite{motwani2020automatically} to
consolidate the SBFL metrics using weighted ranks to identify the top-$\mathpzc{K}$ (35\%) suspicious functions~\cite{lin2010rank}.
We refer to our approach and its output as {coarse-grained fault localization} (CGFL).

\SUBSECTION{\textbf{\AutomatedProgramRepair.}}
Automated program repair (\APR) methods generate patches for defects in software
with minimal or no human
intervention~\cite{Goues_Pradel_Roychoudhury_2019}.
There are many popular methods (see Gazzolla \emph{et al.} for a survey~\cite{Gazzola_Micucci_Mariani_2017}), but most have adopted a \emph{search-based} approach,
defining transformation operators, e.g., different flavors of mutation, to manipulate existing code, and using test suites to validate repair correctness.  Other methods use formal semantics either alone or in combination with mutation-based search.
Recently, machine learning (ML) repair like neural machine translation~\cite{ye2022neural} or large language models~\cite{fu2022vulrepair} have proliferated. 
While most ML models require perfect fault location (source code line or function), they are not currently appropriate for binary repair as they are overwhelmingly trained on source code or natural language.
However, \PRD could be used in conjunction with ML tools, which may require training or fine-tuning with decompiled source code to improve effectiveness.
We evaluate \PRD using two independently developed mutation-based source-code repair tools, 
Prophet~\cite{long2016automatic} and GenProg~\cite{le_goues_systematic_2012,le_goues_genprog_2012} with its deterministic variant ``AE''~\cite{weimer2013leveraging}.

%% file: motivate.tex
\input{prd_binrepared_multifig}

\ifdefined\SHORTFORMMOTIVATINGEXAMPLE
\SECTION{Motivating Example}
\label{sec:motivate}
To illustrate our approach, imagine that we have to fix a buggy binary program, but have no access to the source. We also have 
 some tests that display the bug and expected behavior.
Normally, to statically repair this binary in these conditions, one first root-causes the failure, figures out its proper behavior, transforms this behavior into content compatible to the confines of the original binary, and then rewrites the binary with this content.
\PRD automates this typically manual process by applying source-based APR algorithms to binaries to search for satisfactory patched binaries, retaining these repairs in high-level source. 
\PRD takes as input a set of functions likely associated with the bug (CGFL) and generates a new binary.
To accomplish this, \PRD employs existing decompilers to lift binary functions to high-level source, transforms this into compatible binary content via binary-source interfaces, then modifies these binary functions to adhere, then jump to these interfaces.
By patching source code into existing binaries, \PRD not only simplifies the normally difficult tasks of statically rewriting a binary and generating compatible binary content, but also enables analyses at the source code level for this content, e.g., \APR algorithms.
Altogether, we obtain the ease and benefit of source-level \APR, applied to a binary. 
\else
\begin{lstlisting}[
style=whiteCstyle,
rulesep=1pt,
framextopmargin=1pt,
framexbottommargin=1pt,
xleftmargin=7pt,
captionpos=t,
breakatwhitespace=false,
%belowcaptionskip=0.5em,
%belowskip=0em,
%aboveskip=0em,
boxpos=c,
%float=*,
%floatplacement=t,
label=code:extractcgcsplit,
language=C,
columns=fullflexible,
caption={Decompiled code for \texttt{\small KPRCA\_00018}'s \texttt{\small cgc\_split}, generated by the Hex-Rays decompiler. Low readability of decompiled code does not limit APR tools.\\
%	\vspace{-0.55cm}
}
]
int cgc_split() {
  int v0; int result; char v2; card_t *v3; card_t *v4; 
  squarerabbit_t *split_srabbit; squarerabbit_t *srabbit; 
  int i; i = 0;
  for(srabbit = g_srabbit; srabbit->player_finished && i < cgc_split_len(); srabbit = &split_hand[v0])
      v0 = i++;
  if(!srabbit->double_or_split || !cgc_can_split(srabbit))
      return -1;
  v2 = g_srabbit->split_len;
  g_srabbit->split_len = v2 + 1;/*BUG*/
  if( v2 > 1 ) return -1;       /*BUG*/
\end{lstlisting}\vspace{-0.35cm}
\SECTION{Motivating Example}
\label{sec:motivate}
To motivate and illustrate our approach, consider \texttt{\small KPRCA\_00018 ({Square\_Rabbit})} from the DARPA CGC dataset, a casino-inspired game with an integer overflow vulnerability 
that can crash the program.
Let's assume we need to prevent crashes, but the software is no longer supported, the source is unavailable, and direct binary fault localization and patching are not feasible~\cite{hu2019automatically}. 
However, we have recovered some tests for the buggy binary (\textit{100} functional, one crash-inducing) and have access to a source- and test-based APR tool. Our APR tool does not need source code to be particularly readable, so we rely on decompiled code that recompiles 
and is test-equivalent to the original.  
To accomplish this, we assume Hex-Rays as our off-the-shelf decompiler. 
First, we apply a decompiler to the binary which fails to generate recompilable output for all functions.
Although we cannot fully decompile, we can use PRD to address this. By recovering compatible type definitions and decompiling functions individually, we can generate a large proportion of recompilable (87) and test-equivalent (76) functions from the binary.

\ifdefined\MOTEXTCGFL
\begin{table*}
	\footnotesize
	\centering
	\caption{Square\_Rabbit CGFL: Top 7 Results for five state-of-the-art SBFL metric. We show the first 7 of 43 functions and their suspiciousness scores. Suspiciousness scores closer to 1.0 indicate are more likely to be vulnerable.}
\resizebox{\textwidth}{!}{
	\input{data/cgfl_square_rabbit.tex}

}
\label{table:cgfl_sq}
\end{table*}
\begin{lstlisting}[belowcaptionskip=0em,
belowskip=0em,
aboveskip=0em,
boxpos=c,
float=tp,
floatplacement=tp,
basicstyle=\tiny,
caption={CGFL: ordered output from Rank Aggregation of top-35\% (15) functions},
label=rafl_out]
The optimal list is: 
cgc_remove_card cgc_split cgc_remove_from_blist cgc_stand cgc_discard_hand
cgc_all_hands_busted cgc_get_card cgc_is_player_finished cgc_cardtos
cgc_deal_srabbit cgc_clear_squarerabbit cgc_can_split cgc_print_cards
cgc_print_winner cgc_new_srabbit_game
\end{lstlisting}
\fi

\begin{sloppypar}	
We localize the likely source of the problem to a small set of suspicious functions, typically less than 20, in the binary (fault localization), successfully implicating a set that includes the vulnerable \texttt{\small cgc\_split} 
\ifdefined\MOTEXTCGFL
function (see Table~\ref{table:cgfl_sq} for individual SBFL metrics and 
Listing~\ref{rafl_out} for the CGFL).
\else
function, as the most suspicious (tied in rank 1 with one other function).
\fi

\end{sloppypar}

The next step is generating a patched binary that preserves the expected functionality and repairs the vulnerability.
We decompile the suspicious function set, recover compatible types ({partial decompilation}), and transform them into source that is both recompilable and compatible with binary rewriting via binary-source interfaces.
To verify that the resulting binary is test-equivalent to the original binary, we recompile the decompiled code and generated binary-source interfaces, customize the binary content to construct a new binary (partial recompilation and binary rewriting), and check the result for test equivalence.
Finally, we address the vulnerability by repairing the decompiled source, either manually or automatically with APR, recompiling and customizing the patch to create a repaired binary.

Using our example, Listing~\ref{code:extractcgcsplit} shows the buggy \texttt{cgc\_split} function as decompiled by Hex-Rays (bug appears on lines 10--11: incrementing {\small{\texttt{g\_srabbit->split\_len}}} results in an integer overflow).
Our APR tools successfully identify a developer-equivalent patch (i.e., moving \texttt{\small 10} after \texttt{\small 11}). 
Altogether, we obtain the ease and benefit of source-level \APR, applied to a binary. 
\fi

%% file: prd_binrepared_multifig.tex

\begin{figure}[!b]
	\centering
	\caption{Stages of \PRD and \sysacronym with stage-consistent (color)s. }
\subfloat[{Description of input requirements and high-level goals for \PRD, broken up into stages.
	\label{fig:prd_highlevel}} ] {
	\includegraphics[width=1.0\linewidth]{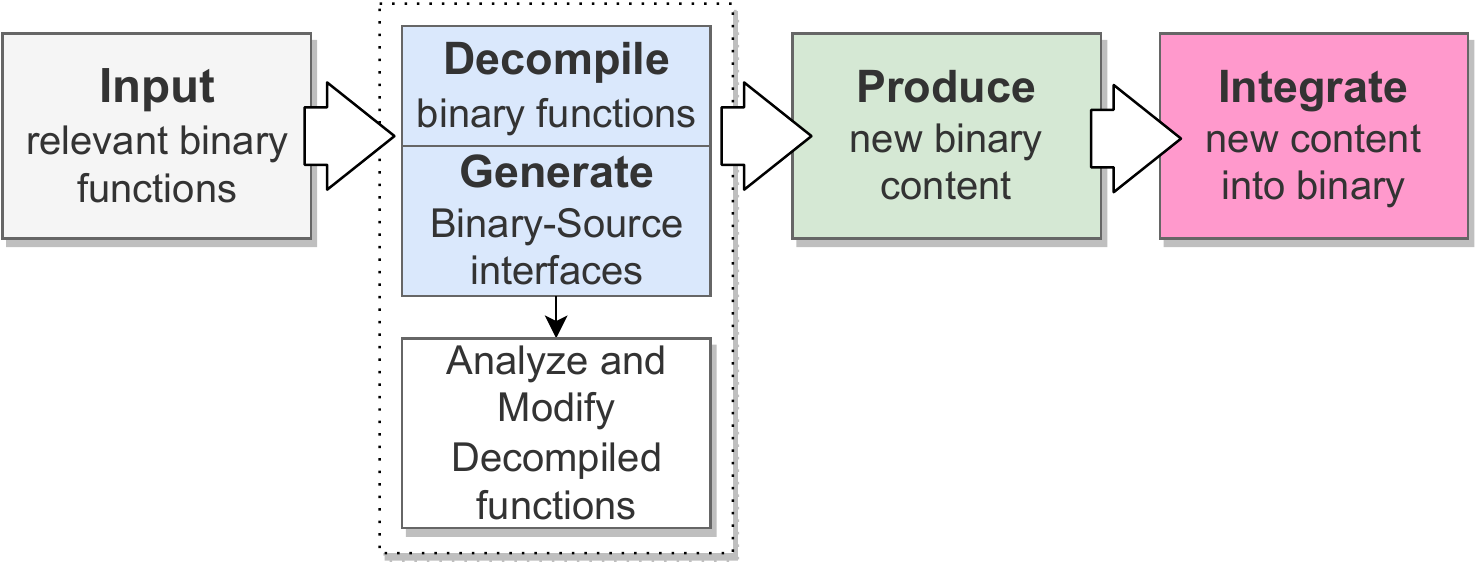}
}
\vspace{-0.3cm}
\hfill
	\subfloat[{The \PRD method: \circled{i}partial decompilation~{(blue)}, \circled{ii}partial recompilation~{(green)}, \circled{iii}binary rewriting {(rose)}.
	 \label{fig:prd}} ] {
		\includegraphics[width=1.0\linewidth]{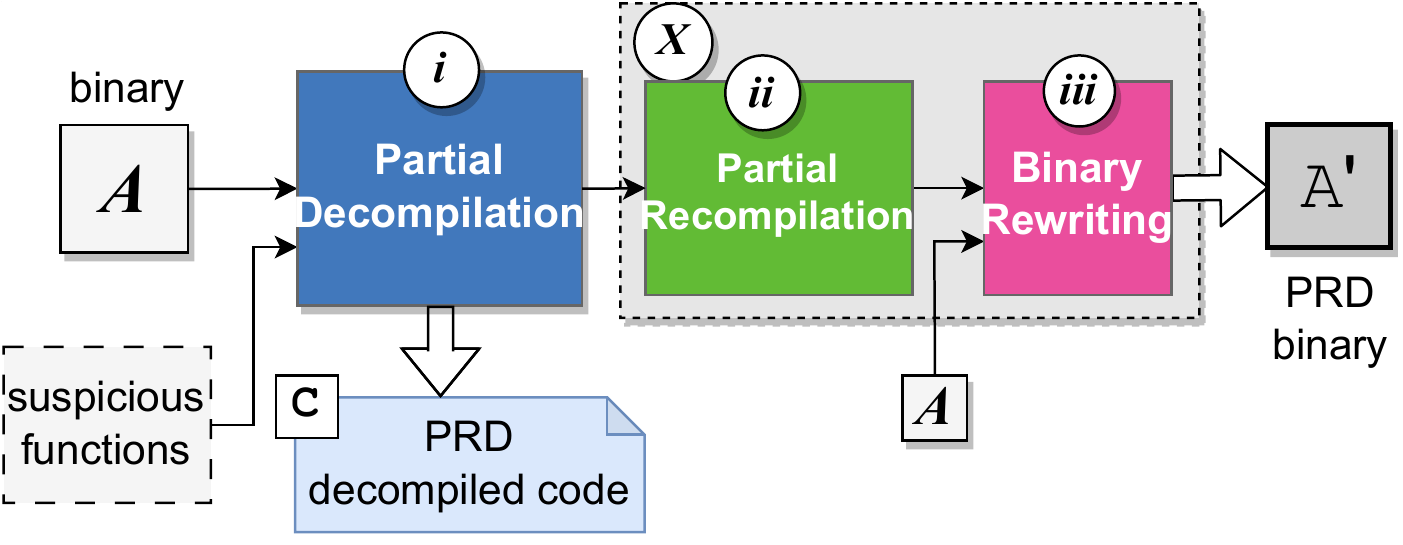}
	}
	\hfill
	\subfloat[{The four stages of \sysacronym: \circled{1}CGFL; \circled{2}partial decompilation; \circled{3}source-level repair ({\APR}\textsubscript{prd}); \circled{4}partial recompilation and binary rewriting. Each stage's output is formatted in same style as its label.}
	\label{fig:apr_prd}]{%
		\includegraphics[width=1.0\linewidth]{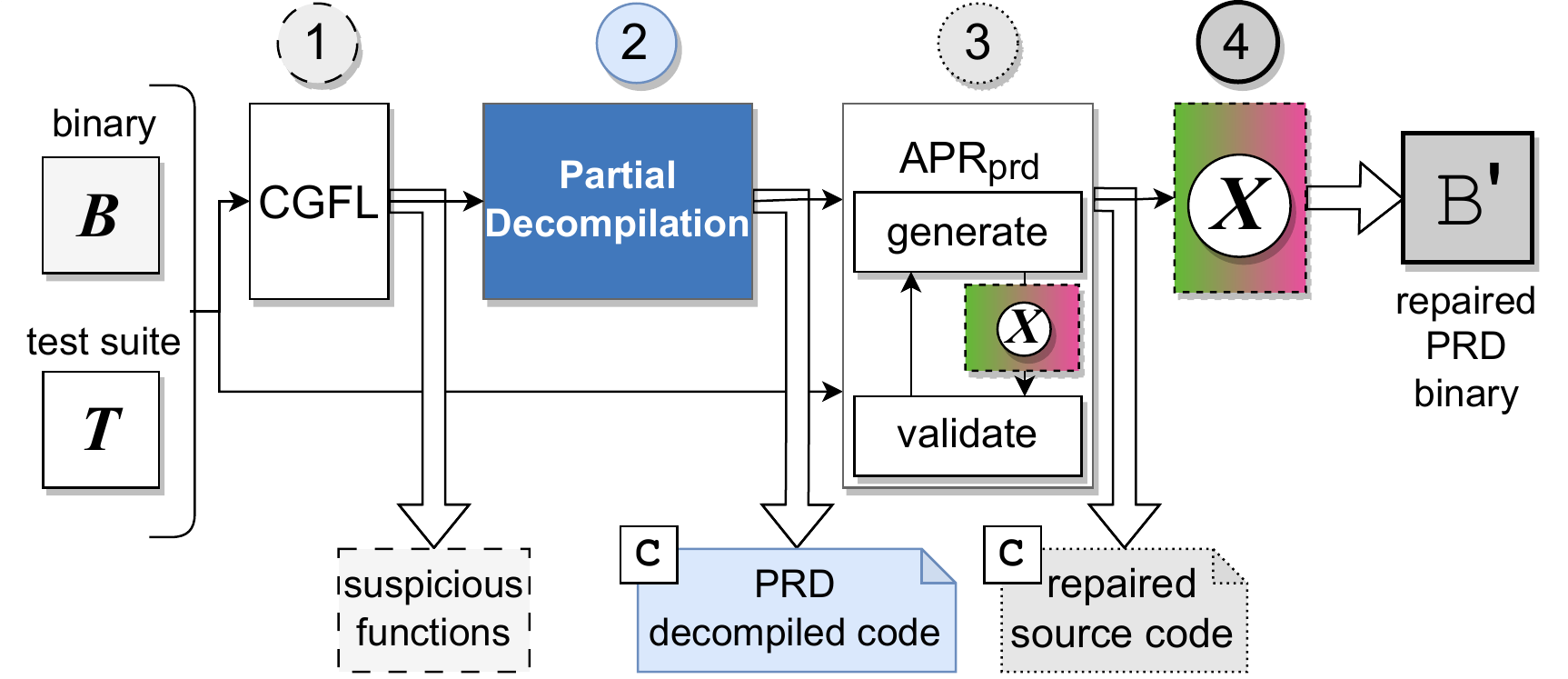}
	}

%
	\vspace{-0.5cm}
	\label{fig:ALLPRD}
\end{figure}

%% file: data/cgfl_square_rabbit.tex
\begin{tabular}{|p{0.2cm}|Q{3.0cm}Q{1.0cm}|Q{4.0cm}Q{1.0cm}|Q{3.0cm}Q{1.0cm}|Q{3.5cm}Q{1.0cm}|Q{3.0cm}Q{1.0cm}|}
\hline
	 N&\multicolumn{2}{P{4.0cm}|}{Tarantula}
	 &\multicolumn{2}{P{5.0cm}|}{ochiai}
	 &\multicolumn{2}{P{4.0cm}|}{$op^2$}
	 &\multicolumn{2}{P{4.5cm}|}{Barinel}
	 &\multicolumn{2}{P{4.0cm}|}{Dstar(star=2)}    \\
\hline
	 0&cgc\_remove\_card&1.000&cgc\_remove\_card& 1.000&cgc\_split     &  1.000&cgc\_split       &1.000&cgc\_remove\_card&1.000 \\
	 1&cgc\_split    &   1.000&cgc\_split      & 1.000&cgc\_remove\_card& 1.000&cgc\_remove\_card& 1.000&cgc\_split &      1.000\\
	 2&cgc\_remove\_from\_blist& 0.885&cgc\_remove\_from\_blist &0.718&cgc\_remove\_from\_blist &0.875&cgc\_remove\_from\_blist &0.516&cgc\_remove\_from\_blist &0.500\\
	 3&cgc\_stand    &   0.858&cgc\_stand   &    0.676&cgc\_stand    &   0.842&cgc\_stand       &0.457&cgc\_stand &      0.441\\
	 4&cgc\_is\_player\_finished &0.852&cgc\_print\_winner &0.667&cgc\_can\_split  & 0.833&cgc\_calc\_score & 0.444&cgc\_dealer\_hit&  0.429\\
	 5&cgc\_calc\_score & 0.852&cgc\_check\_player\_squarerabbit &0.667&cgc\_get\_card    &0.833&cgc\_shuffle\_deck\_if\_needed &0.444&cgc\_print\_cards &0.429\\
	 6&cgc\_dealer\_hit & 0.852&cgc\_new\_srabbit\_game &0.667&cgc\_print\_cards &0.833&cgc\_discard\_hand &0.444&cgc\_cardtos &    0.429\\
\hline
\end{tabular}

%% file: technical.tex
\label{sec:technical}
\SECTION{\PartiallyRecompilableDecompilation}
\label{sec:prd_tech}
Our functional prototype contribution, \PRD enables source-level analysis techniques and vulnerability mitigation, even when only the binary is available.
Unlike reassemblers or binary recompilers which operate on low-level software abstractions, \PRD enables analyses to be performed at the source code level. 
While this is a strength of \PRD, it comes at the cost of additional requirements, as well as analytical and engineering difficulties.
Since compilers do not support combining new content with the non-object binary content, our methodology has to effectively perform linking and locating with all new content. 
Additionally, any resulting binary must retain the same executional qualities as the original, such as the ability to call and use external and local symbols regardless of their binding.

\PRD takes as input a small set of binary functions (identified by
fault localization) and
consists of three interdependent custom stages that uphold our requirements: partial decompilation, partial recompilation, and binary rewriting, detailed in Figure~\ref{fig:ALLPRD}.
The output is a single binary, the \PRD binary, composed of the original binary, binary-source interfaces, and decompiled function. 

While it does not recompile the original source code, \PRD does recompile the source code it generates. 
We consider \emph{\PRD decompiled code} to be this generated source code, i.e., decompiled functions and binary-source interfaces.
These binary-source interfaces allow decompiled code to execute original binary content, and when used in conjunction with customized detours,
 allow original binary content to execute decompiled code. 
When \PRD decompiled code is compiled, we consider it to be the \PRD recompiled content.
As a baseline, \PRD succeeds when its output binary is test-equivalent to the original binary.

Our prototype implementation operates on x86 statically- and dynamically-linked Linux ELF executables,
compatible with System V ABI and can be extended to x86-64 and stripped binaries with engineering effort.
The rest of this section discusses key \PRD concepts: partial decompilation, partial recompilation, binary rewriting, and fault localization.


\SUBSECTION{Partial decompilation}
\label{sec:decompilation_tech}
Decompilation output can be produced by automated decompilers, human experts, or a combination (e.g., experts editing decompiler output).
While \textit{full decompilation} refers to the complete decompilation of a binary,
we say a decompilation is \emph{partial} when only certain functions in the binary are decompiled and compatible types are recovered.
This design choice mitigates some of the weaknesses and limitations of current binary decompilers and enables source-only analyses (such as APR) on binary code before binary decompilation achieves perfection.
The vulnerable functions, inferred types, and other dependencies are extracted from the binary and decompiled to source. 
To support later \PRD stages, additional analyses are necessary to generate compatible binary-source interfaces.

\SUBSUBSECTION{Requirements}
\label{sec:decomp_req}
Because partial decompilation is required to generate source that is compatible with the later PRD stages, 
we discuss the 
constraints necessary to use the decompiled source.

\begin{figure}[]
	\centering
	\includegraphics[width=0.5\textwidth]{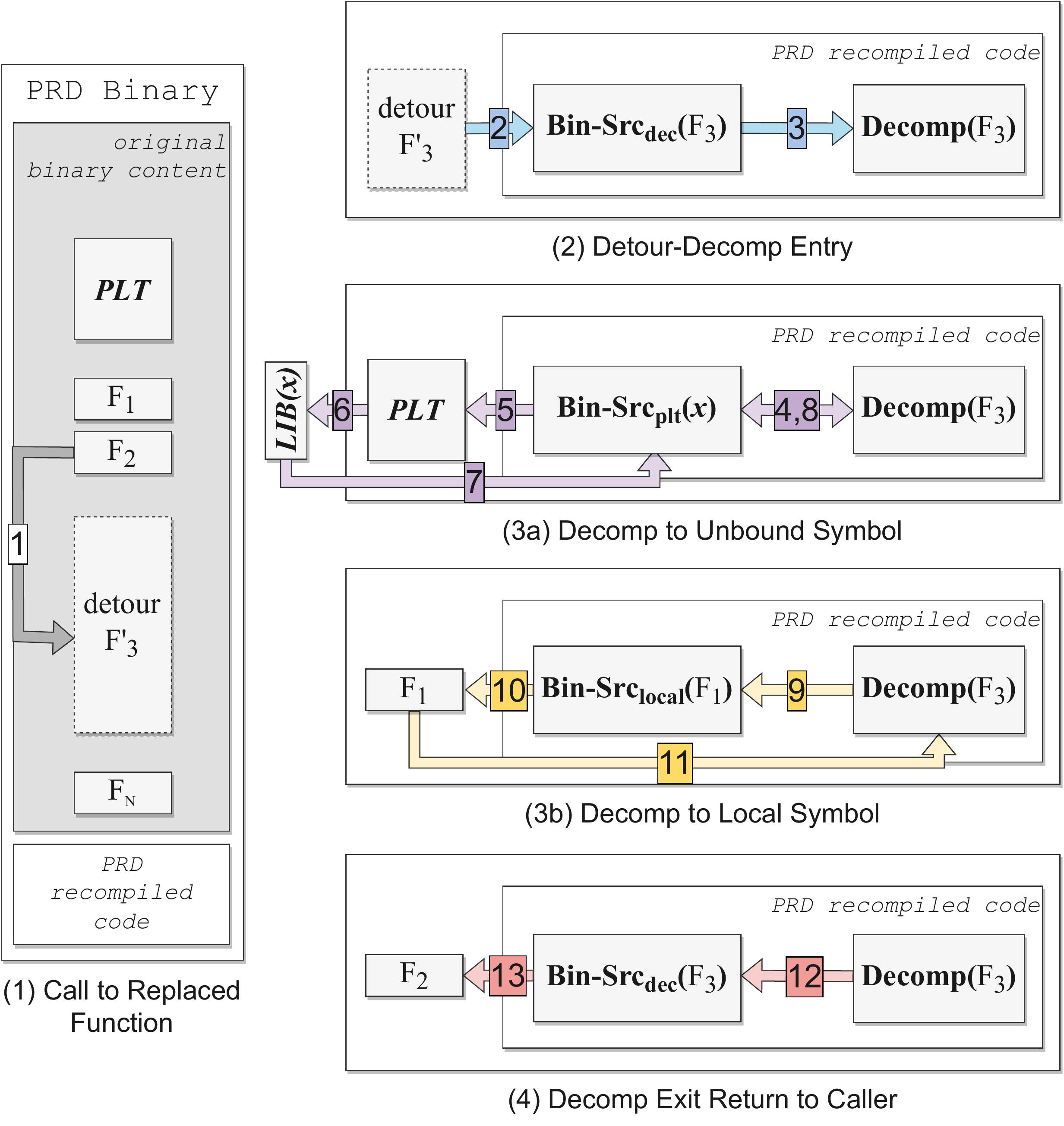}
	\caption{Program Execution Flows for \PRD replacement function for \emph{F\textsubscript{3}} with each fundamental type of execution flow between original and new binary content outlined by way of numbered arrows. 
		From the original binary, \emph{F\textsubscript{1...N}} refer to its binary functions and  
		\emph{PLT} refers to its Procedure Linkage Table (ELF).
		New binary content, i.e., \PRD recompiled code, consists of
		\emph{Decomp(\emph{F\textsubscript{3}})}, the decompiled content for function \emph{F\textsubscript{3}}, and \emph{Bin-Src\textsubscript{...}}, binary-source interfaces. Subscripts \emph{dec} specify the detour interface for a decompiled function, \emph{plt} the unbound symbol interface for a dynamically resolved symbol, and \emph{local} the interface for a local symbol, whose location is resolved. 
	 }
	 \label{fig:prd_execution}
\end{figure}

\SUBSUBSUBSECTION{{\PRD recompiled content and its constraints}}
\label{sec:decomp_constraints}
Because customized detours connect the original binary to decompiled content and  all linking and locating is done by \PRD, the standard interpreter initializations like constructors, symbol resolutions and relocations, are bypassed for the added decompiled content.
To accommodate this, we ensure that the \PRD recompiled content fulfills these key requirements:
(\textbf{r1.1}) does not require the interpreter
and
(\textbf{r1.2}) references global, local, and external symbols in original binary content.
In Figure~\ref{fig:prd_execution}, we outline the requisite high-level execution flows that \PRD enables between original binary content and decompiled content. 
We focus on the decompiled content and its
dependencies (symbols).

\SUBSUBSUBSECTION{{Binary-Source Interfaces}}
To satisfy these constraints, we analyze both binary and decompiled content, then generate two complementary binary-source interfaces: unbound symbol and detour interfaces.  
These allow decompiled code to reference symbols (like callbacks) from the original binary regardless of these symbols' binding state. We illustrate examples of these interfaces in Listing~\ref{code:exampledetour}.
As the entry to dynamically-linked functions,
the unbound symbol interface wraps a callback with code that allows dynamic linking to behave as if it were being called from the original binary. For the ELF binaries, this interface manages the procedure linkage table, PLT. 
Similarly, as the entry point to recompiled, decompiled function content, the detour interface manages both PLT and all required symbols used by this content. 
Together, they ensure the resulting recompiled content is consistent to the original. 
\begin{sloppypar}
	The detour interface adds required symbols to the original function prototype, shown in Listing~\ref{code:exampledetour} with added \texttt{\small void*} parameters that harbor values for \texttt{\small ebx} register and three references: \texttt{\small cgc\_receive},  \texttt{\small cgc\_memcpy}, and \texttt{\small cgc\_calloc}.
	Because \texttt{\small cgc\_memcpy} and \texttt{\small cgc\_calloc} are local symbols at calculable locations in the binary, they can be invoked as callbacks from the decompiled code. However, \texttt{\small cgc\_receive} is a symbol whose binding state is indeterminable by decompiled code during runtime, we create an unbound symbol interface to allow decompiled code to interface with its PLT entry.
This means that the original binary's function call and the detour interface have diverged; ergo, a lone jump instruction, \texttt{e9}, will not suffice.  
\end{sloppypar}

Divergence from the original function call to the detour interface has these implications: (\textbf{i1}) our detour interface is not compatible with the original function call; (\textbf{i2}) the stack state is not consistent upon return from decompiled function; (\textbf{i3}) added references, $r$, incur a byte-cost, $c=7r+4$,  which may overflow the original function.
In Section~\ref{sec:binary-manipulation}, we explain how \PRD's binary rewriting phase handles (i1).  
\PRD decompilation addresses (i2) by inserting stack-correcting
inline assembly before the detour interface's return.
Note on (i3), when multiple functions are decompiled, the minimum set of required references for any entry function is the union of its calltree's references.

This approach satisfies our requirements: r1.1 (the binary-source manages references) and r1.2 (the decompiled code uses the original binary's symbols).

\begin{lstlisting}[
caption={Example \textbf{binary-source interfaces} generated by PRD: unbound symbol \texttt{\scriptsize (cgc\_receive)} and detour \texttt{\scriptsize (det\_cgc\_read\_line)} interfaces, with local symbol callbacks \texttt{\scriptsize (cgc\_calloc)} and \texttt{\scriptsize (cgc\_memcpy)}. \\
},
style=whiteCstyle,
rulesep=1pt,
framextopmargin=1pt,
framexbottommargin=1pt,
xleftmargin=7pt,
multicols=2,
captionpos=t,
%abovecaptionskip=5pt,
%belowcaptionskip=30pt,
%belowskip=5pt,
%aboveskip=12pt,
boxpos=b,
float=*,
%floatplacement=b,
label=code:exampledetour,
language=C,
columns=fullflexible
]
// EBX mechanism needed to interface with original binary's PLT
unsigned int origPLT_EBX = NULL;

// required symbol-1 : typedef function ptr 
typedef int (*pcgc_receive)(int s_0, int s_1, int s_2, int s_3);
pcgc_receive z__cgc_receive = NULL; 

// required symbol-1 : Unbound Symbol interface 
int cgc_receive (int s_0,int s_1,int s_2,int s_3) {
    pcgc_receive lcgc_receive = z__cgc_receive;
    int ret;
    unsigned int localEBX;
    unsigned int localorigPLT_EBX = origPLT_EBX;
    asm ( "movl %[LOCALEBX],%%ebx\n\t"
          "movl %%ebx,%[PLT_EBX]\n\t"
          :[LOCALEBX] "=r"(localEBX)
          :[PLT_EBX] "r"(localorigPLT_EBX)
          : "%ebx");
    ret = lcgc_receive(s_0,s_1,s_2,s_3);
    asm ( "movl %%ebx,%[LOCALEBX]\n\t"
          :[LOCALEBX]"=r"(localEBX));
    return ret;
}

// required symbol-2 : typedef function ptr
typedef void * (*pcgc_memcpy)(void *, const void *, cgc_size_t);
pcgc_memcpy cgc_memcpy = NULL;

// required symbol-3 : typedef function ptr
typedef void * (*pcgc_calloc)(cgc_size_t);
pcgc_calloc cgc_calloc = NULL;

// Decompiled Function Declaration
cgc_ssize_t  cgc_read_line(int fd, char **buf);

// Detour Interface 
cgc_ssize_t  det_cgc_read_line( 
    // binary-source interface references
    void* EBX, void* mycgc_receive, void* mycgc_calloc, void* mycgc_memcpy,
    // parameters from Decompiled Function
    int fd, char * * buf )
{
    cgc_ssize_t retValue;
    origPLT_EBX = (unsigned int) EBX;
    z__cgc_receive = (pcgc_receive)(mycgc_receive);
    cgc_calloc = (pcgc_calloc)(mycgc_calloc);
    cgc_memcpy = (pcgc_memcpy)(mycgc_memcpy);

    retValue = cgc_read_line( fd, buf);
    asm( "mov    -0xc(%ebp),%eax\n\t"
         "mov    -0x4(%ebp),%ebx\n\t" "nop\n\t"
         "add    $0x14,%esp\n\t" "nop\n\t"
         "pop %ebx\n\t"
         "pop %ebp\n\t"
         "pop %ecx\n\t"
         "add $0xc,%esp\n\t"
         "push %ecx\n\t"
         "ret\n\t"
    );
    return retValue;
}

\end{lstlisting}

\SUBSUBSECTION{Decompilation.} Here, we outline our use of decompilers and their output.
\SUBSUBSUBSECTION{Function-specific Decompilation}
\label{sec:func_decomp}
For PRD to succeed at binary repair, not all binary functions need to be decompiled.
Instead, we apply decompilers to a sufficient subset, like CGFL (Section~\ref{sec:cgfl}).
This decompiled output is left intact, minus a small set of decompiler-specific keyword substitutions that are necessary for APR and recompilation (e.g., replace \texttt{\small DWORD} with \texttt{\small unsigned int}).

\SUBSUBSUBSECTION{Compatible Type Recovery}
\label{sec:type_recovery}
Similarly, 
it is not necessary to recover the exact types from the binary, but instead \emph{compatible types} are required.
For example, a struct \texttt{\small foo} may contain many fields with different types, but only one of the fields (e.g., \texttt{\small foo.bar} where \texttt{\small bar} is an unsigned int) is used in a decompiled function.
In partial decompilation, we only need to recover the offset and infer the type of \texttt{\small bar} for the decompiled function to be suitable for PRD.
This significantly reduces the requirement of complete and sound type inference on binary code, an open research problem.
To accomplish compatible type recovery, we leverage decompilers' type inference to reconstruct the necessary compatible types from the binary.  
Notably, although decompilers can fail to recover all types, \PRD can succeed if only referenced compatible types are defined.
Since types may be nested, PRD decompilation resolves a definition order for the compatible type definitions to ensure recompilability.

\SUBSUBSECTION{Implementation}
\label{sec:decomp_impl}
\begin{sloppypar}
Our prototype primarily uses Hex-Rays Decompiler. 
While Hex-Rays generates an initial
decompilation, our custom IDAPython script obtains
corresponding local type and function declarations.
Note, to aid \texttt{\small C++}-decompilation, we use Ghidra and Hex-Rays.
\PRD substitutes common primitives such as \texttt{\_DWORD} and any Hex-Rays-specific definitions, resolves a
definition order for types, generates the required binary-source interfaces and resolves the minimum set of required symbols. 
These rule-based transformations are a best-effort heuristic to produce informative decompilation.
To reorient the stack to support detouring with references, we analyze the initial recompiled \PRD source, generate inline assembly that reconstructs the stack on detour exit, then insert this assembly snippet in the detour interface (see lines 34-42 of Listing~\ref{code:exampledetour}).
\end{sloppypar}
	
\SUBSECTION{Partial Recompilation}
\label{sec:tech_partialrecomp}
\emph{\PartialRecompilation} must recompile
high-level source code to ensure its correct execution in context of
the patched binary, as well as support r1.1 (Section~\ref{sec:decomp_req}).
Because our \binarymanipulation strategy appends new binary content,
it must operate even if the memory address of new
content is not known in advance.
To satisfy these requirements, partial recompilation creates {position-independent}, statically linked content.
While position-independent code (PIC) is ubiquitous, support for both static linking and PIC in a single shared object is recent, i.e., \texttt{\small-shared -static-pie} (supported in GNU 8.4.0 and later), and not behaviorally consistent across all compilers.
We generate our object with these flags and custom linker script, placing all sections in a single segment.
Although GCC-compatible, we use \emph{dietlibc}, a small-footprint libc, to support APR profiling.
Our approach supports executional requirements and r1.1.

\SUBSECTION{Binary Rewriting}
\label{sec:binary-manipulation}
Binary rewriting composes the original binary and \PRD-decompiled source into a single binary that executes correctly.
Our prototype analyzes, extracts, adds, and manipulates binary content to satisfy our requirements, r1.1-2.
Satisfying (i1) (Section~\ref{sec:decomp_constraints}), we change the effective function call to align with the detour interface, before inserting a jump to it.
To accomplish this, we analyze the binary content to generate instructions for each required reference, such that the result adheres to calling convention. 
We implemented our prototype in Python with LIEF~\cite{thomas2017lief}.

\SUBSECTION{Coarse-Grained Fault Localization (CGFL)}
\label{sec:cgfl}
Our approach focuses on the minimum requirement, i.e., 
identifying a small set of functions to decompile and patch, ideally including the vulnerable function. 
By applying an SBFL variant, i.e., correlating binary function spectra to their execution, a set of suspicious functions can be identified for decompilation.
For simplicity, we refer to our FL approach and its output as CGFL.

\SUBSUBSECTION{Requirements}
\label{subsec:cgfl}
Although the three stages of \PRD operate on a small number of \emph{implicated functions}, 
if the vulnerable functions are not implicated, \sysacronym cannot succeed. 
Because the binary setting provides different assumptions from most FL methods (Section~\ref{subsec:fl}), 
this results in five CGFL requirements: 
(\textbf{r2.1}) does not require source code or the ability to recompile, 
(\textbf{r2.2}) prioritizes functions for decompilation, 
(\textbf{r2.3}) minimizes run-time overhead,
(\textbf{r2.4}) avoids functions that cannot support our detouring,
and ultimately 
(\textbf{r2.5}) identifies vulnerable functions.  
After, finer-grained FL can pinpoint suspicious statements.

\SUBSUBSECTION{Implementation}
\label{sec:cgfl_imp}
CGFL uses \emph{Valgrind}'s~\cite{nethercote2007valgrind} \emph{callgrind} to trace the program under different unit tests~(satisfying r2.1--r2.3). 
Valgrind can efficiently provide function level trace, allowing us to identify function-level spectra. 
To address requirement r2.4 (detour support), we implemented a screening algorithm that reduces probability of overruns and, for statically-linked programs, culls standard library functions.
To avoid detour overruns, we set a minimum function size of 45 bytes, which supports at least six references per detour.  
This eliminates very small functions with large calltrees (Section~\ref{sec:decomp_constraints}).

For each qualified function, we calculate suspiciousness scores using five state-of-the-art SBFL metrics (Tarantula~\cite{jones2002visualization}, Ochiai~\cite{hutchins1994experiments}, o$p^2$~\cite{naish2011model}, Barinel~\cite{abreu2009spectrum}, $D^2$~\cite{wong2013dstar}). 
Using matrices generated from this content, RankAggreg identifies the top-$\mathpzc{K}$ suspicious functions~\cite{lin2010rank}
(satisfying r2.5: This implicates vulnerable functions in the
top 35\% over 92\% of the time).
This generates CGFL: a prioritized, consolidated list of suspicious functions from our function SBFL metrics.

\ifdefined\MOTEXTCGFL
We illustrate CGFL using the \texttt{\small {Square\_Rabbit}} example (Section~\ref{sec:motivate}). 
Although callgrind identifies more than 278 calls, only 43 are functions that correspond to local symbols. 
Focusing on these 43 functions, we generate suspiciousness scores for each SBFL metric
 (see Table~\ref{table:cgfl_sq}).
We generate the new two matrices, names and weights, across each SBFL metric.
RankAggreg generates a list of the top-$\mathpzc{K}=15$ (35\%) suspicious functions from Square\_Rabbit, shown in Listing~\ref{rafl_out}. 
\else

\fi

%% file: dataset.tex
\SECTION{Experimental Setup}

\noindent{}
Our evaluation of \PRD and \sysacronym includes evaluations of underlying assumptions, as well as studies on end-to-end fully automated scenarios. 
 Additionally, we consider two real-world case studies (Section~\ref{sec:eval-cve}). 
 Specifically, our evaluation addresses the following research questions:
\begin{enumerate}[label=\textbf{RQ\arabic*.},leftmargin=1cm,topsep=2pt]
  \item Without any restrictions, how often is decompiled code recompilable? 
  \item Is decompiled code behaviorally consistent to original binary functions? 
  \item Does CGFL identify function(s) relevant to the vulnerability?
  \item How effective is \sysacronym at mitigating vulnerabilities?
  \item[{\textbf{Case}}]\textbf{{Study} : Generality.} Does \PRD generalize to real-world vulnerabilities, other languages, and performance constraints?
\end{enumerate}
\noindent
Next, we describe our datasets and experimental setup.
\SUBSECTION{Benchmark Datasets}
\label{sec:dataset}

\begin{sloppypar}
Outlined in Table~\ref{table:datasets}, our four benchmark datasets are:
the DARPA Cyber Grand Challenge \texttt{\small C} binaries (\texttt{CGC-C})
and \texttt{\small C++} binaries (\texttt{CGC-C++});
Rode0day 19.11 (\texttt{Rode0day}); 
and vulnerable, real-world programs (\texttt{CVE Case Study}). 
\end{sloppypar}
\SUBSUBSECTION{CGC-C and CGC-C++}
\ifdefined\SHORTFORMDATASETS
The DARPA 2016 Cyber Grand Challenge ({CGC}) provides a dataset of challenge binaries ({CBs}), each containing realistic vulnerabilities, and a testing framework.
We derived our CGC datasets from a {Linux} variant,
\emph{trailofbits}~\cite{trailofbits/cb-multios_2020}, verified these {CBs} using a robust variant of the testing environment.
This identified 110 valid {CGC CBs} (100-\texttt{\small CGC-C}) and 10-\texttt{\small CGC-C++}), those with at least 9 passing positive and one negative test, i.e., defect scenarios.
For our \texttt{\small CGC-C} evaluations, we consider \emph{target}s for each CB, a set of functions for decompilation, totaling 190 as some binaries have multiple vulnerable functions.
\else
At the binary level, 
the CGC dataset is a compelling example of the use cases we envision, where it is crucial to address a vulnerability quickly, and support or source may not be available.
The DARPA 2016 Cyber Grand Challenge ({CGC}) provides a dataset of challenge binaries ({CBs}) that each contain realistic vulnerabilities, and testing framework.
We derived our CGC datasets from a {Linux} variant,
\emph{trailofbits}~\cite{trailofbits/cb-multios_2020}.  We verified all 
{CBs} using a robust variant of the testing environment and found that 110 {CGC CBs} (100
from \texttt{\small C} source, 10 from \texttt{\small C++}) had at least one failing negative and 9 passing 
positive tests, see Table~\ref{table:datasets}. As each binary may have more than one vulnerability, we identified 157 \emph{scenarios}, i.e., the defect scenario to repair, for \texttt{\small CGC-C} and 10 for \texttt{\small CGC-C++}.
For our \texttt{\small CGC-C} evaluations, we consider an additional feature for each {CB}: \emph{target}, a set of functions targeted for decompilation, totaling 190 applicable targets (some binaries have multiple vulnerable functions).
\fi 

\SUBSUBSECTION{Rode0day}
\label{sec:dataset_rode0day}
\ifdefined\SHORTFORMDATASETS
Rode0day 19.11 inserts hundreds of bugs into Linux binaries, each a collection of stripped binaries and example test inputs. We recompile to ensure that symbol information exists in the binary (required by PRD) and generate unit tests from provided inputs (see Table~\ref{table:datasets}).
For our evaluations, we consider both functions and the injected bugs.
\else
Rode0day is a competition series which focuses on finding bugs in binaries. Rode0day inserts hundreds of bugs into Linux binaries, each a collection of stripped binaries and example test inputs. After a competition concludes, Rode0day releases an archive with this content, crash-inducing inputs found by participants, and source code annotated with crash ids. 

We derived our dataset from the archived release of Rode0day 19.11, recompiled to ensure that symbol information exists in the binary (required by PRD) and generated unit tests from provided inputs (see Table~\ref{table:datasets}).
For our evaluations, we consider both functions and the injected bugs.
\fi 

\SUBSUBSECTION{Real-World Case Study}
Finally, we also consider two real-world programs, \texttt{\small podofopdfinfo} (\emph{PoDoFo-0.9.7}~\cite{podofo}) and \texttt{\small jhead} (\emph{jhead-3.0.6}~\cite{jhead}), not specifically
curated for automated repair or binary analysis, which have associated public security vulnerabilities (CVEs).
These programs fit the minimum \PRD requirement---the vulnerable methods are local symbols.
\begin{table}[b]
	\footnotesize
	\centering
	\setlength\tabcolsep{3.5pt}
	\caption{Evaluation Dataset features: release information, number of binaries, number of defect scenarios, average and minimum number of behavioral (pos) tests, compiler-toolchain and source language, and average lines of code (LOC) for each dataset. 
	}	
	\resizebox{0.5\textwidth}{0.06\textheight}{\tiny\ttfamily
		\begin{tabular}{||O{1.0cm}O{1.25cm}Q{0.4cm}Q{0.8cm}Q{1.0cm}O{0.8cm}Q{0.7cm}||}
			\hline
			 & Release & \# bins & \# defect scenarios&avg. pos  (min)& Compiler- Lang& avg. LOC \\
			 \hline

			\multicolumn{7}{||O{5.95cm}||}{Curated Dataset}\\
			\rowcolor{LGray}
			CGC-C & trailofbits & 100 & 157 & 104.0~~(9)  & GNU-C & 37,670\\
			CGC-C++ & trailofbits & 10 & 10 & 89.8~(13) & GNU-C++ & 1,138\\
			\rowcolor{LGray}
			Rode0day & 19.11 & 4 & 927 & 1.5~~(1) & GNU-C & 84,725\\		
			\hline

			\multicolumn{7}{||O{5.95cm}||}{Real-World Case Study}\\
			\rowcolor{LGray}
			2021-30472 & PoDoFo-0.9.7 & 1 & 6 & 36 & GNU-C++ & 47,414\\
			2021-3496 & jhead-3.0.6 & 1 & 1 & 68 & Clang-C & 4,203\\
			\hline
		\end{tabular}
	}
	\label{table:datasets}
\end{table}

\SUBSECTION{External Tools}
\label{sec:eval-compatibility}
\PRD uses Hex-Rays IDA Pro 7.5 SP2
(Hex-Rays), GCC 8.4.1, and dietlibc. When handling \texttt{\small C++},
we augment Hex-Rays with the Ghidra SRE Public 10.0.1.

For \sysacronym, we use Valgrind for CGFL and apply two APR tools: Prophet (v0.1, Clang-based) and GenProg (v3.2, CIL-based). 
We used Prophet's default search algorithm with its \emph{profile} localizer and three of GenProg's: 
default genetic algorithm (``GA''), 
a deterministic search focused more on
static analysis (``AE'')~\cite{weimer2013leveraging},
and a ``single-edit'' repair search.
When evaluating program variants, GenProg replaces standard compilation
with \PRD tools.
Prophet replaces the compiler with custom scripts, relying on environmental variables and dynamic libraries (we make each compatible to \PRD).
Ultimately, \PRD operates seamlessly.

%% file: evaluation.tex
\SECTION{Empirical Evaluation}
\label{sec:evaluation}

In this section, we present the results of our evaluation and explain how they answer our research questions. 
The first three questions evaluate underlying assumptions, the remaining evaluate our automated approach.

\SUBSECTION{\small RQ1: How often is decompiled code recompilable?}
\label{sec:eval-decompilation}

To answer this research question, we studied \texttt{\small CGC-C}, \texttt{\small CGC-C++}, and \texttt{\small Rode0day} datasets and place no restrictions on compilation or on the source code grammar.
Our evaluation focuses these decompilation features: type recovery, decompilation, basic recompilability, and binary reconstruction.
Our type recovery evaluation considers each binary independently. 
If all specified types are not recovered, then type inference fails for that binary.
Decompilation and recompilability evaluations consider each binary function independently and asks how many could be decompiled by Hex-Rays and recompiled.
If Hex-Rays fails, then decompilation fails for that function.
Basic recompilation per function fails if any dependency is not fully defined, i.e., compatible type recovery, prototype recovery, and decompilation.
Although we evaluate basic recompilation with raw decompiler output for function prototype recovery and function content, because this raw output rarely compiles, we use our \PRD-transformed types. 
To assess the impact of type recovery failures on basic recompilation, we include the Incomplete Typing failure rate.
\begin{table}[tb]
	\scriptsize
	\centering
	\caption{Evaluation results for RQ1 and RQ2. For RQ1, success rates for Type Recovery, Decompilation, and Basic Recompilation are shown; failure rates from Incomplete Typing. For RQ2, success rates for PRD Recompilation and Test-equivalency are shown.
	Only Type Recovery is tabulated per binary, the rest are per function. \vspace{-0.2cm}}
	\resizebox{\linewidth}{0.06\textheight}{\tiny\ttfamily
	\input{data/partrecomp_eval.tex}
	}
	\label{table:partrecomp}
	\vspace{-0.55cm}
\end{table}

Our results, outlined in Table~\ref{table:partrecomp}), demonstrate that decompilation succeeds more often than not, but basic recompilation is impacted by type recovery failures.  
When the decompiler recovered all required types,
70-89\% of functions succeed at basic recompilation.
Although we evaluated binary reconstruction for each binary, only two \texttt{\small CGC-C} binaries (Palindrome, Palindrome2) are fully reconstructable, i.e., all its functions succeed at basic recompilation. 

\FRAMED{We find that only 1.7\% of binaries are fully recompilable, but 70-89\% of individually decompiled functions are recompilable when typing succeeds. This strongly supports our insight to use \emph{partial}, instead of full decompilation.}

\SUBSECTION{RQ2: Is decompiled code behaviorally consistent to original binary functions? }
\label{sec:eval-prd}
To answer this question, we assessed partially decompiled content on our \texttt{\small CGC-C}, \texttt{\small CGC-C++}, and \texttt{\small Rode0day} datasets: (general) how often can partially decompiled content result in a test-equivalent binary? 
In both, we applied \PRD decompilation to individual functions, extending basic recompilability from Section~\ref{sec:eval-decompilation} with \PRD transformations, e.g., generating \PRD binary-source interfaces to create \PRD decompiled code.
Using this output, we then leveraged \PRD's recompilation and binary rewriting methods to generate a new \PRD binary. 
Finally, each \PRD binary is evaluated using the corresponding binary's test-cases (test-equivalency).
For behavioral tests, any disparity between the \PRD binary and original is considered a failure, while success of proof-of-vulnerability tests includes mitigation of the vulnerability in addition to consistent behavior.

Our results, shown in Table~\ref{table:partrecomp}), demonstrate that recompilation of \PRD decompiled code performs consistently to basic recompilation from Section~\ref{sec:eval-decompilation} (i.e., \PRD decompilation transformations do not introduce errors and are impacted similarly by incomplete types) and that resulting \PRD binaries are often test-equivalent to their original, exceeding 92\%.  
We also evaluated Binary reconstruction and behavior for each binary, like earlier, only two \texttt{\small CGC-C} binaries (Palindrome, Palindrome2) could be fully reconstructed and could generate test-equivalent \PRD binaries for all functions.

\FRAMED{These results show that when decompilation succeeds, \PRD produces patched binaries that are test-equivalent to the original. \PRD provides a solid foundation for	source-level transformations.}

\SUBSECTION{RQ3: Does CGFL identify function(s) relevant to the vulnerability?}
\label{sec:eval-cgfl}
As most FL methods use source-based or expensive dynamic instrumentation methods, we explore the viability of our CGFL implementation to identify suspicious functions from the vulnerable binary.
After confirming compatibility with our implementation, we used our three dataset's test-cases as stimuli and annotations as ground truth (i.e., vulnerable functions that should be implicated). 

For all binaries in our datasets, our results show that the CGFL output contains at least one ground-truth function for 95 of 100 \texttt{\small CGC-C}, 8 of 10 \texttt{\small CGC-C++}, and 196 of 206 \texttt{\small Rode0day}.
When accounting for all ground-truths, we see similar success: 74 \texttt{\small CGC-C}, 7 \texttt{\small CGC-C++}, and 196 \texttt{\small Rode0day}, which succeeds at 95\% despite having few tests (Section~\ref{sec:dataset_rode0day}).

While CGFL succeeds more than 92\% with our criteria, when CGFL failed to identify a vulnerable function, we observed three failure types that can be readily explained or mitigated.
First, 14 binaries (4 \texttt{\small CGC}/10 \texttt{\small Rode0day}) did not exercise any vulnerable function in any negative test (f.1).  
Second, 10 were in the first three ranks, but ties impacted their selection (f.2), a common failure in SBFL metrics.
Third, 1 buggily reimplemented a libc function (f.3).
Although (f.1) cannot be addressed by SBFL or by APR, (f.2) can readily mitigated by adding better test content or increasing the size of $\mathpzc{K}$, per~Section~\ref{subsec:cgfl}. 
Finally, (f.3) is a result of our simple heuristic that screens out system functions, statically located in the binary, which could be replaced with a more sophisticated screening process.
\FRAMED{Even when few tests are available, our coarse-grained fault localization method works well in practice. CGFL contains the relevant function in 291/316 cases.}
\begin{table}[]
	\scriptsize
	\centering
	\caption{APR Comparison: Full-source (baseline) vs. \sysacronym with \PRD decompiled code (PRD).
		We report the number of scenarios that produced a plausible
		mitigation (\textit{mitigated}), the \textit{total} number that the APR tool successfully launched its serach, as well as the number which the tool \textit{completed} its search within 8 hours.\vspace{-0.2cm}
	}
	\resizebox{0.5\textwidth}{0.035\textheight}{
		\tiny\ttfamily
	\input{data/binrepared_apr_compare.tex}
	}
	\label{table:prd_e2e_apr_tab}
\vspace{-0.55cm}
\end{table}

\SUBSECTION{\small RQ4: How effective is \sysacronym at mitigating vulnerabilities?}
\label{sec:eval-ee-apr}
\ifdefined\TWOAPRSCENARIOS
Our evaluation consists of two scenarios: (plausibility) can APR tools leverage partially decompiled content to find repairs?; (content) does the form and quality of decompiled content impact APR tools? 
\subsubsection{Plausibility.}
\else
\fi
Because \APR tools do not always succeed, we compared the success rate for \APR tools applied to the actual source
code of the binary (baseline) to the success rate for the same \APR tool applied to the \PRD decompiled code.
Our primary evaluation features an end-to-end use of \sysacronym, as shown in Figure~\ref{fig:apr_prd}).
CGFL implicates a subset of the binary, then partially decompiled to create \PRD decompiled code, the input to the \APR algorithm. 
To identify repairs, the \APR algorithm uses \PRD tools to apply source-level patches to the original binary and test content to evaluate the resulting \PRD binary's behavior. 
Unlike \sysacronym, the baseline has access to the entire source.

Using the 157 defect scenarios from the \texttt{\small CGC-C} dataset, we assessed baseline and \PRD-enabled APR algorithms, limiting each run to 8 hours, outlining results in Table~\ref{table:prd_e2e_apr_tab}.
Our primary result shows that \PRD-enabled repair, operating only on binaries,
performs as well as and sometimes better than full-source repair:
\PRD-supported algorithms find 51--69 plausible patches, while
the full-source baselines find 32--57. 
Prophet performs slightly better with access to the original source (57 vs.\ 52) while the 
GenProg variants perform better in the \PRD setting (51--69 vs.\ 32--48).
Collectively, our PRD-supported APR tools mitigated 85 of the 148 scenarios (including 20 that the winning DARBA CGC cyber-reasoning system, Mayhem, did not patch).
Overall, we find that the success rate for repairs operating on binaries via \sysacronym
as good as the same tools operating on full-source ($p<0.0004$, proportions z-test).
Because these \APR algorithms rely on the global availability of source code, 
\sysacronym should be less likely to succeed due to the reduction in available source. 
Our results are surprising.
\SUBSUBSECTION{Case Studies: Repair Quality} 
APR tools can find repairs that overfit the test
suite~\cite{Smith_Barr_LeGoues_Brun_2015,Le_Thung_Lo_Goues_2018} without
addressing the root cause of the problem.  
We do not improve or worsen that orthogonal concern here, instead find
that \sysacronym inherits the repair quality of its underlying APR method.
In our use case, quickly disrupting an exploit is valuable, even if the
repair is not general.  
We focus on GenProg results for simplicity, Prophet results are similar.

First, in cases where GenProg failed to produce a mitigation,
the required edit was usually out of scope, like changes to struct fields or unused variables.
Other failures involved special constant values or comparators, known
tool weakness (cf.~\cite{Oliveira_Souza_Goues_Camilo-Junior_2018}).
Second, we consider the mitigations that it did find.  
As multiple solutions were found, we randomly sampled 5--10\% and examined their \texttt{\small C} representations (recall from
Section~\ref{sec:technical}, \sysacronym retains source-level patches, facilitating such analyses).
We do not find significant difference in the rate
of overfitting between the full-source \APR baseline and \sysacronym. 
As a case study, we describe one example that mitigates the vulnerability but does not address it generally (overfits), and then a second example that corrects the
defect in a more general way.

\SUBSUBSUBSECTION{Lower-Quality}
For \texttt{\footnotesize KPRCA\_00013}'s
first vulnerability, GenProg mutation \texttt{\footnotesize a(1178,1056)} passes all tests and successfully mitigates it.  In this edit, the ``\texttt{\footnotesize (}''
character is pushed onto the operator stack in a loop, but the next
iteration flags an error since the top of the stack is
``\texttt{\footnotesize (}''.  Although this patch does not address the
official 
vulnerabilities for this CB, it does mitigate this particular
exploit, preventing control of the next heap block's heap metadata.

\SUBSUBSUBSECTION{Higher-Quality}
\texttt{\footnotesize NRFIN\_00076}'s first
defect simulates a vulnerability introduced when a
programmer commits unfinished code.  The program incorrectly increments
\texttt{\footnotesize *results} in a frequent function, 
leading to the use of an invalid pointer. 
The mutation found, \texttt{\footnotesize d(4)}, correctly deletes the problematic code, eliminating
the pointer issues.

\FRAMED{Our primary \sysacronym result shows that in an end-to-end scenario, using \PRD to apply a source-level \APR tool to binaries produces results that \emph{are consistent with and sometimes better than} using those same techniques on the corresponding source. This is true both in terms of the rate at which vulnerabilities are mitigated and in terms of repair quality (overfitting).}
\ifdefined\TWOAPRSCENARIOS

\begin{table*}[t]
	\scriptsize
	\centering
	\caption{Experimental \sysacronym  evaluation on 30 scenarios.
		\textit{LOC} refers to the lines-of-code for DARPA CGC \emph{CB} source code and \emph{PRD} decompiled code;
		\textit{fns}: number of decompiled functions; \textit{{AST}}: size of decompiled CIL-AST. Our results for APR tools using full-source are indicated by \emph{baseline}, with APR\textsubscript{prd} for \emph{PRD} (decompiled code) and \emph{exact} (exact decompilation).
		We use \texttt{-} for decompiler errors; \emph{$\times$} APR tool issues;
		\emph{$+$} APR profiling failures that expose vulnerability;
		\emph{\checkmark} successful repairs; \emph{\textbar} practice binaries.\vspace{-0.2cm}
	}
	\resizebox{0.8\textwidth}{0.16\textheight}{
	\scriptsize\ttfamily
	\input{data/genprog_prd_cgc_table.tex}
	}
	\label{table:genprog_cgc_tab}
	\vspace{-0.45cm}
\end{table*}

\SUBSUBSECTION{Study of Input Content on APR Effectiveness.} 
\label{sec:decomp_eval_ideal}
To assess the impact of our informative
decompilation approach, 
we studied a random sample of 21 CGC CBs and their 30 scenarios (Table~\ref{table:genprog_cgc_tab}).
We consider access to the full source of the program (\emph{baseline}),
the \PRD-provided decompiled source for implicated methods (\emph{PRD}), and
a bound in which the CGC-provided source is used for the same implicated methods
(\emph{exact}), i.e., the exact source function replaces the decompiled function.
We observe that GenProg and Prophet separately produce 28 candidate patches with \emph{baseline},
18 with \emph{PRD}, and 34 with \emph{exact}
decompilation. 
This echoes the RQ1 finding that state-of-the-art decompilation tools have room to improve in end-to-end usage scenarios. 

In addition, Table~\ref{table:genprog_cgc_tab}'s LOC values offer a partial explanation for
how decompilation can perform comparably to full source: the decompilation
applies only to the implicated functions (reducing effective LOC by 95\%), while the full source provides the entire source code to the \APR tool. 
This supports our insight that CGFL, if accurate at implicating a relevant
subset, can help downstream stages (such as decompilation or
program analysis/transformation).
\FRAMED{Our secondary \sysacronym result shows that decompilation can strongly impact quality, but
	current decompilers' weaknesses are mitigated by fault localization filtering.
}
\fi

\subsection{Case Study: Generality}
\label{sec:generality}
Our previous results establish some generality, but for a binary-facing technique it is also important to consider multiple languages, real-world programs and defects, and execution overhead.

\SUBSUBSECTION{Application to \texttt{C++}.}
\label{sec:eval-ee-cpp}
We have shown that \PRD can partially decompile and recompile binaries produced from \texttt{\small C++} (see Table~\ref{table:partrecomp}) for individual functions.
Because we cannot directly evaluate baseline \APR algorithms on \texttt{\small C++} (an unsupported language), in this end-to-end assessment, we focus on evaluating further compatibility with \sysacronym, i.e., multiple \texttt{\small C++} functions. 
Using the \texttt{\small CGC-C++}, we evaluate \PRD's applicability to \sysacronym specifically using each binary's CGFL as the decompile set. 
For each binary, we applied \PRD to its CGFL, totaling 149 functions (includes decompiled \texttt{\small C++} class methods).
Although we generate test-equivalent \PRD binaries for 7 of the 10 binaries,
recompilation failures (94) dominate test (10) and decompilation (9) failures, like our Section~\ref{sec:eval-prd} results.
\noindent
These results indicate that the \PRD framework is not only extensible to
\texttt{\small C++} binaries, but can potentially allow repairing
\texttt{\small C}-like binaries with \texttt{\small C}-based APR tool.
Because APR tools overwhelmingly favor other languages over C++, our results improve tool applicability.  

\SUBSUBSECTION{Real-world Vulnerabilities.}
\label{sec:eval-cve}
We detail \PRD and CGFL effectiveness for two real-world programs that contain known vulnerabilities (see Table~\ref{table:datasets}).

\begin{sloppypar}
\noindent{\small \textit{CVE-2021-30472}:}
 For tests, we used the developer's recommended example PDFs and the CVE input, \texttt{\footnotesize bug4} for \texttt{\footnotesize podofopdfinfo}.
	This exploit hits a vulnerability in {\texttt{\footnotesize PdfEncryptMD5Base::ComputeEncryptionKey}} (\texttt{\footnotesize f1}) before 
	{\texttt{\footnotesize PdfEncryptMD5Base::ComputeOwnerKey}}~(\texttt{\footnotesize f2}) vulnerability.
	We applied our CGFL approach (Section~\ref{subsec:cgfl}).
	All SBFL metrics identified \texttt{\footnotesize f1} in rank 1 (10 ties) of 265 with \texttt{\footnotesize f2} at rank 147---echoing the bug~precedence.
We successfully applied \PRD to both methods.
Because Hex-Rays incorrectly lifted the {stack canary} using x86-64 content, we had to manually correct this with the equivalent x86 inline assembly. 
Using the GCC-G++ compiled binary and this content, \PRD successfully generated a test-equivalent binary for \texttt{\footnotesize f2}, 
while \texttt{\footnotesize f1} improved on test-equivalency, mitigating the vulnerability. Specifically, the decompiler generated a large number of local variables for \texttt{\footnotesize f1}, changing the stack such that the original vulnerability was no longer intact. 

\noindent{\small \textit{CVE-2021-3496}:}
We applied our same methods to \texttt{\footnotesize jhead}.
CGFL implicated the function, \texttt{\footnotesize ProcessMakerNote}, 
as rank 3 of 42. Notably, it inlines the reported buggy function 
\texttt{\footnotesize ProcessCanonMakerNoteDir}.
\PRD successfully generated a test-equivalent binary from a Clang binary.
\end{sloppypar}

By manually applying the necessary bug fixes (less than 3 lines were changed), we produce \PRD-binaries that both pass all tests, resolving the CVE-reported bugs for both jhead and podofopdfinfo, successfully applying \PRD to two real-world issues.

\SUBSUBSECTION{Performance Analysis.}
Using \emph{perf stat}, we consider run-time and compile-time performance for \PRD binaries.
For run time, we compared 100 \PRD binaries to their counterparts over 5,163 tests and found that the difference in performance was not statistically significant {\small (user: $p < 0.970$; system: $p < 0.277$, two-tail t-test)}.
\noindent For compile time, an overhead relevant to \APR
algorithms, we sampled 25 CGC CBs and respective single-detour \PRD binaries,
generating each 25 times.
We found that generating a \PRD binary is statistically less expensive than compiling the binary from source {\small (user:$p<0.0$; system: $p<8.4845e^{-192}$, two-tail t-test)}.
These results imply that \PRD does not induce a performance overhead on \APR tools.

\FRAMED{We show that \PRD can be applied to \texttt{\small C++} binaries and real-world CVEs. Our rate of producing test-equivalent binaries for our \texttt{\small C++} benchmark, 24.2\%, is comparable to our rate for \texttt{\small C}.  We find that both CGFL and \PRD are successful on 2 of 2 real-world CVEs. We find negligible overhead for producing and running \PRD binaries.}

%% file: data/partrecomp_eval.tex
\begin{tabular}{||O{0.75cm}|P{0.75cm}P{0.75cm}P{0.75cm}|P{1.0cm}|P{1.0cm}|P{1.0cm}P{1.0cm}||}
		\cline{1-8}
	& \multicolumn{4}{P{3.25cm}|}{RQ1: Partial decompilation} & \multicolumn{3}{P{3.0cm}||}{RQ2: Partial \mbox{recompilation}} \\
	\cline{2-4}\cline{5-5}\cline{6-8}	
	Dataset&\multicolumn{3}{c|}{Success} & Failure & Failure & \multicolumn{2}{c||}{Success}\\
	\cline{1-5}\cline{6-8}
	 & Type \mbox{Recovery} & \mbox{Decompile} & Basic \mbox{Recompile} & Incomplete Typing  & Incomplete Typing & PRD \mbox{Recompile} & Test Equivalent \\
	
	\rowcolor{LGray}
	CGC-C & 54.0\% & 99.9\% & 78.7\% & 54.6\% & 56.6\% &89.9\% & 92.1\% \\
	\rowcolor{LGray}
	&  54/100 & 7060/7067 & 5553/7060 & 823/1507 & 823/1455 &5605/6237 & 5161/5605 \\
	{CGC-C++} & 0.0\% & 87.8\% & 52.6\% & 80.3\%  &78.7\% & 70.1\% & 97.0\% \\
	& 0/10 & 1099/1251 & 379/1099  & 578/720 & 578/734& 365/521 & 355/366 \\
	\rowcolor{LGray}	
	{Rode0days} & 50.0\% & 100.0\% & 30.5\% & 81.5\% &82.5\% & 71.8\% & 96.4\% \\
	\rowcolor{LGray}
	& 2/4 & 3297/3297 & 620/3297 & 1865/2289 & 1880/2280& 1016/1416 & 979/1016 \\
	\hline\hline
	{Total} & 49.1\% & 98.6\% & 60.6\% & 72.3\% & 73.4\% & 85.5\% & 93.0\% \\
	\cline{1-8}
	\multicolumn{8}{c}{}
\end{tabular}

%% file: data/binrepared_apr_compare.tex
\begin{tabular}{@{}O{0.90cm}*{4}{|P{0.7cm}P{0.5cm}}||@{}}
	\cline{2-9}
	&\multicolumn{2}{P{1.2cm}|}{``AE''}&
	\multicolumn{2}{P{1.2cm}|}{GenProg \mbox{``single edit''}}&
	\multicolumn{2}{P{1.2cm}|}{GenProg\newline\mbox{``GA''}}&
	\multicolumn{2}{P{1.2cm}||}{Prophet}\\
	\cline{1-9}
    \multicolumn{1}{||O{0.90cm}|}{}& baseline & PRD & baseline & PRD & baseline & PRD & baseline & PRD  \\
	\rowcolor{LGray}
	\multicolumn{1}{||O{0.90cm}|}{Total} & 137 & 157 & 129 & 157 & 94 & 157 & 79 & 157 \\
	\multicolumn{1}{||O{0.90cm}|}{Completed} & 122 & 137 & 113 & 129 & 67 & 94 & 79 & 79 \\
	\hline\hline
	\multicolumn{1}{||O{0.90cm}|}{Mitigated} & 45 & \textbf{69} & 48 & \textbf{69} & 32 & \textbf{51} & \textbf{57} & 52 \\
	\cline{1-9}
\end{tabular}

%% file: data/genprog_prd_cgc_table.tex
\begin{tabular}{||O{1.5cm}Q{0.3cm}P{0.45cm}|P{0.6cm}|Q{0.9cm}Q{0.7cm}|P{0.4cm}P{0.7cm}*{3}{|P{0.75cm}}*{3}{|P{0.75cm}}|P{0.7cm}||}
\hline
        DARPA & fn & pov & \# & \multicolumn{2}{P{1.6cm}|}{LOC} & \multicolumn{2}{P{1.1cm}|}{PRD} & \multicolumn{3}{P{2.25cm}|}{Prophet}& \multicolumn{3}{P{2.25cm}|}{GenProg} & \\
\cline{9-11}\cline{12-14}
       CGC CB & id & id & tests & CB & PRD & fns & AST & base line & PRD & exact & base line & PRD & exact & Mayhem \\
\hline
             KPRCA\_00013 &     0 &  1 &         44 &   2,360 &          1,917 &                7 &         349 &        \checkmark &            \checkmark &        \checkmark &                     \checkmark &            \checkmark &        \checkmark &            \\
             KPRCA\_00013 &     0 &  2 &   $\cdots$   & $\cdots$ &        $\cdots$ &                  &             &        \checkmark &            \checkmark &        \checkmark &                     \checkmark &            \checkmark &        \checkmark &            \\
\rowcolor{Gray}
 NRFIN\_00041 &     0 &  1 &         84 &   1,138 &            510 &                8 &         207 &      &          &      &                     \checkmark &         $\times$ &      &  \checkmark \\
             CROMU\_00027 &     4 &  2 &        200 &   7,502 &             83 &                3 &         129 &    $\times$ &          &      &                   &          &      &            \\
             CROMU\_00027 &     4 &  3 &   $\cdots$   & $\cdots$ &        $\cdots$ &                  &             &    $\times$ &          &      &                     \checkmark &          &      &            \\
             CROMU\_00027 &     4 &  4 &   $\cdots$   & $\cdots$ &        $\cdots$ &                  &             &    $\times$ &          &      &                   &          &      &            \\
             CROMU\_00027 &     4 &  5 &   $\cdots$   & $\cdots$ &        $\cdots$ &                  &             &    $\times$ &          &      &                   &          &      &            \\
\rowcolor{Gray}
 KPRCA\_00010 &     1 &  1 &         99 &   1,920 &            744 &               14 &         185 &        \checkmark &            \checkmark &      &                     \checkmark &          &        \checkmark &  \checkmark \\
             KPRCA\_00009 &     0 &  1 &        100 &   8,397 &            107 &                3 &          11 &        \checkmark &            \checkmark &        \checkmark &                     \checkmark &         $\times$ &        \checkmark &            \\
             KPRCA\_00009 &     0 &  2 &  $\cdots$    & $\cdots$ &        $\cdots$ &                  &             &        \checkmark &            \checkmark &        \checkmark &                     \checkmark &         $\times$ &        \checkmark &            \\
             KPRCA\_00009 &     0 &  3 &  $\cdots$    & $\cdots$ &        $\cdots$ &                  &             &        \checkmark &            \checkmark &        \checkmark &                     \checkmark &         $\times$ &        \checkmark &            \\
\rowcolor{Gray}
 CROMU\_00001 &     0 &  1 &        101 &  43,394 &            126 &                3 &          41 &      &  - &        \checkmark &  $\times$ &  - &        \checkmark &  \checkmark \\
             YAN01\_00010 &     0 &  1 &        100 &     228 &            123 &                3 &          14 &      &  - &      &                     \checkmark &  - &        \checkmark &  \checkmark \\
\rowcolor{Gray}
 NRFIN\_00075 &     0 &  1 &         87 &     882 &            178 &                1 &         892 &        \checkmark &  - &      \checkmark &                 $\times$ &  - &      \checkmark &            \\
             CROMU\_00033 &     0 &  1 &        100 &     988 &            105 &                1 &          53 &      &            \checkmark &        \checkmark &                   &            \checkmark &        \checkmark &            \\
\rowcolor{Gray}
 CROMU\_00032 &     0 &  1 &        101 &   1,151 &            313 &                3 &         134 &      &  - &        \checkmark &                   &  - &      &  \checkmark \\
             NRFIN\_00035 &     0 &  1 &        100 &    6755 &            140 &                1 &         757 &      &          &      &                     \checkmark &         $\times$ &      &            \\
\rowcolor{Gray}
\rowcolor{Gray}
 KPRCA\_00017 &     0 &  1 &         29 &   1,225 &            162 &                3 &          50 &      &          &      &                   &        $+$ &      &            \\
\rowcolor{Gray}
 KPRCA\_00017 &     0 &  2 &  $\cdots$    & $\cdots$ &        $\cdots$ &                  &             &        \checkmark &            \checkmark &        \checkmark &                   &        $+$ &        \checkmark &            \\
             KPRCA\_00019 &     0 &  1 &        100 &   1,399 &            216 &                3 &          49 &        \checkmark &  - &        \checkmark &                   $\times$ &  - &        \checkmark &            \\
             KPRCA\_00019 &     0 &  2 &  $\cdots$    & $\cdots$ &        $\cdots$ &                  &             &        \checkmark &  - &        \checkmark &                   $\times$ &  - &        \checkmark &            \\
\rowcolor{Gray}
 CROMU\_00037 &     0 &  1 &        101 &  52,123 &            124 &                3 &          26 &        \checkmark &            \checkmark &        \checkmark &       $\times$ &            \checkmark &        \checkmark &  \checkmark \\
             CADET\_00001 &     0 &  1 &         99 &     208 &             97 &                4 &          33 &      &          &      &                     \checkmark &          &      &      \textbar\\
\rowcolor{Gray}
 KPRCA\_00073 &     0 &  1 &         93 &   1,918 &            129 &                3 &          17 &      &            \checkmark &        \checkmark &                   &         $\times$ &      &            \\
             KPRCA\_00060 &     0 &  1 &         51 &   1,051 &            244 &                3 &          74 &        \checkmark &            \checkmark &        \checkmark &                   &         $\times$ &        \checkmark &      \textbar\\
\rowcolor{Gray}
 NRFIN\_00076 &     0 &  1 &        100 &   1,811 &             52 &                3 &           8 &      &  - &        \checkmark &                 $\times$ &  - &        \checkmark &      \textbar\\
             KPRCA\_00007 &     0 &  1 &         32 &   1,631 &            111 &                3 &          16 &        \checkmark &            \checkmark &        \checkmark &                     \checkmark &          &      &            \\
\rowcolor{Gray}
\rowcolor{Gray}
 NRFIN\_00020 &     0 &  1 &         92 &     720 &            139 &                6 &          57 &      &          &      &                   &         $\times$ &      &            \\
\rowcolor{Gray}
 NRFIN\_00020 &     0 &  2 &  $\cdots$    & $\cdots$ &        $\cdots$ &                  &             &      &          &      &                     \checkmark &         $\times$ &      &            \\
             CROMU\_00010 &     0 &  1 &        100 &  15,270 &            101 &                1 &           2 &        \checkmark &            \checkmark &        \checkmark &                     \checkmark &            \checkmark &      &  \checkmark \\
\hline\hline
       Repairs/Total &       &      &            &         &                &                  &             &               14/26 & \textbf{13}/23 & 18/30 & 14/24 & \textbf{5}/11 & 16/30 &            \\
\hline
\end{tabular}

%% file: discussion.tex
\SECTION{Discussion}
\label{sec:discussion}
With an initial analysis identifying a relevant set of functions which is then decompiled, \PRD
enables manual and automated mitigation of binary-level exploits. 
\PRD produces a new binary that executes new content instead of the original.

If a complete test suite is not available, then this relevant subset identification may be impacted. However, our CGFL evaluation of Rode0days indicated a 95\% success rate despite a limited test suite, i.e., one behavioral and one exploit.

The current limitations of decompilers are addressed by identifying a small function set, which increases decompilation accuracy and subsequent recompilability.
\ifdefined\TWOAPRSCENARIOS
On average, this approach reduced the lines-of-code (LOC) by an average of 95\% from original to decompiled source (Section~\ref{sec:decomp_eval_ideal}). 
\fi
It is interesting that our \sysacronym results with this reduction are consistent with full-source \APR. 
We leave to future research what role restricting the program search space to a functional subset plays in APR tool efficiency. 

While decompiled source is less readable than the original, it is better than assembly.
This is not an impediment for \APR, as evidenced by the similar
success rates for full-source vs. \sysacronym repairs. 
We can extend \texttt{\small C}-based APR tools to \texttt{\small C++} binaries by decompiling into \texttt{\small C}-like source.
\SUBSECTION{PRD Limitations and caveats.}
\sysacronym and \PRD are not applicable for tools that use whole-program analysis (such as the use of symbolic execution by Angelix~\cite{mechtaev2016angelix}) or interpreted languages. 
While our implementation focuses on 32b ELF and System-V, with engineering effort, \PRD is compatible with other binary formats, ABIs
, and stripped binaries, assuming calling conventions are upheld and decompiler support.
\PRD \emph{is} compatible with ASLR binaries. 
We do not handle self-modifying or self-checking binaries.
The most serious challenge to \sysacronym arises from limitations of current decompilers. When we fail to generate test-equivalent \PRD binaries, decompilation caused most failures.  
Our transformations are limited to currently known and mitigated decompiler weaknesses.
Our evaluations indicate there is some brittleness, particularly with compiler-instrumented binaries and type recovery. 
Our motivating example (Section~\ref{sec:motivate})
and the end-to-end repair results (Section~\ref{sec:eval-ee-apr}) demonstrate the full pipeline with Hex-Rays.

\SUBSECTION{Decompilation failures.}
Although decompilation techniques have achieved impressive results, modern decompilers still struggle
to generate satisfactory results when the binary (1) is compiled from a non-\texttt{\small C} language, (2) contains manual assembly code, or (3) contains self-modifying or obfuscated code.  
Improvements in decompilation enhance \PRD's generality and quality. 

\SUBSECTION{Unsound decompilation results.}
Decompilers do not always generate decompiled code that preserves the semantics of the original binary.
While determining the equivalence of two arbitrary binaries is undecidable in general, this problem can be addressed by requiring byte-level equivalence between the original binary code and the recompiled (but unpatched) code~\cite{Schulte_Ruchti_Noonan_Ciarletta_Loginov_2018}. 
Akin to Equivalence Modulo Input testing~\cite{le2014compiler}, we validate the generated binary against existing test cases, i.e., test-equivalency,
assuming adequate coverage.  

%% file: related.tex
\label{sec:related_work}
\SECTION{Related Work}

Our approach and evaluation rely on recent progress in three major research areas: binary patching and rewriting, APR, and binary code decompilation.

\SUBSECTION{Binary Patching and Rewriting}
There are dynamic and static binary rewriting techniques.
Dynamic binary rewriting or dynamic binary instrumentation 
inserts user code at specified binary locations at runtime, e.g., Pin~\cite{luk2005pin}, Valgrind~\cite{nethercote2007valgrind}, and 
DynamoRIO~\cite{bruening2013building}.
These techniques can introduce prohibitively high overhead and are unused in production binary patching.
Static binary rewriting techniques, like Egalito~\cite{williams2020egalito} and LIEF~\cite{thomas2017lief}, perform code transformation and relocation.
They have much lower runtime overhead than dynamic methods and suit generic tasks like binary patching and control-flow integrity enforcement.
Ramblr~\cite{wang2017} and ddisasm~\cite{flores-montoya2020} convert binary code into assembly that can later be reassembled into a new binary.
E9Patch performs in-binary byte editing in AMD64 binaries to allow insertion of a few chunks of code~\cite{Duck_Gao_Roychoudhury_2020}.
CGC finalists used either in-place binary editing or reassembly to apply patches~\cite{shoshitaishvili2018,Nguyen-Tuong_Melski_Davidson_Co_Hawkins_Hiser_Morris_Nguyen_Rizzi_2018,Avgerinos_Brumley_Davis_Goulden_Nighswander_Rebert_Williamson_2018}.
Unlike \PRD, no existing solutions transform binary code to high-level source.

\SUBSECTION{Automated Program Repair} 
Three recent surveys~\cite{Monperrus_2018a,Gazzola_Micucci_Mariani_2017,Goues_Pradel_Roychoudhury_2019} review more than a decade of this work. 
Most \APR techniques work on source code instead of binary content.
Some exceptions are Schulte \emph{et al.}'s~\cite{schulte2013automated,Schulte_Forrest_Weimer_2010,Schulte_Weimer_Forrest_2015} executable work and Orlov and Sipper's early work~\cite{Orlov_Sipper_2009,Orlov_2017} on Java bytecode.
Since Java bytecode is interpreted by the Java Virtual Machine, ergo incompatible, we omit its discussion. 
Similarly, Angelix~\cite{mechtaev2016angelix}'s symbolic execution is incompatible. 
Tools that assume perfect FL are incompatible~\cite{lutellier_coconut_2020}.
Closely related tools, e.g., RSRepair~\cite{Qi_Mao_Lei_Dai_Wang_2014a}, Kali~\cite{Qi_Long_Achour_Rinard_2015}, and SPR~\cite{Long_Rinard_2015b} would require similar modifications to those we made for our tested tools. 
Other tools, like CodePhage~\cite{Sidiroglou-Douskos_Lahtinen_Long_Rinard_2015}, CodeCarbonCopy~\cite{Sidiroglou-Douskos_Lahtinen_Eden_Long_Rinard_2017} 
are compatible with \PRD but would require more modifications.
OSSPatcher~\cite{Duan_Bijlani_Ji_Alrawi_Xiong_Ike_Saltaformaggio_Lee_2019} targets third-party, open-source libraries for automatic binary patching, but requires both source and source-based patches. 
Current ML-based repair methods, like CoCoNut~\cite{lutellier_coconut_2020} or VulRepair~\cite{fu2022vulrepair}, are not compatible with binary repair, as they not only operate on source code, but also rely on perfect fault localization, i.e., the buggy location is annotated in the input (usually the source code line or context surrounding it).  
However, any ML tool could easily leverage \PRD for both input and binary repair. 

\SUBSECTION{Binary Code Decompilation}
The quality of binary code decompilation relies on advances in binary code extraction, (control flow) structural analysis, and type inference.
Binary code extraction on non-obfuscated binaries is equivalent to control flow graph recovery, where state-of-the-art approaches work in a compiler-, platform-, and architecture-agnostic manner with high precision~\cite{andriesse2017compiler,qiao2017function,flores-montoya2020}.
Structural analysis has progressed significantly:
Schwartz \emph{et al.} reduced the number of \texttt{goto} statements~\cite{schwartz2013};
Yakdan \emph{et al.} proposed pattern-independent control-flow structuring to eliminate \texttt{goto} statements and improve readability~\cite{yakdan2015}.
Decompilers often use static analyses or type inference due to their intrinsic requirement in code coverage (e.g.,~\cite{lee2011,noonan2016,xu2017learning,maier2019}).
Rapid progress in decompilation has enabled the \emph{recompilation} of decompiled code---deemed impossible by most researchers until recently. 
Liu \emph{et al.} show that the output of modern C decompilers is generally recompilable~\cite{Liu_Wang_2020}, when grammar and types are restricted. 
Similarly, Harrand \emph{et al.} present a method that mitigates Java decompiler failures by merging outputs~\cite{harrand2020java}.
Both confirm that decompilers make mistakes and may generate incorrect output.

%% file: conclusion.tex
\SECTION{Conclusion}
\label{sec:conclusion}
Security-critical vulnerabilities that arise after software is deployed must be addressed quickly, even when recompilation is not possible.
Further, 15--25\% of sampled post-release operating system
bug fixes are reported to have end-user visible impacts such as information
corruption~\cite{fixes-become-bugs}.
We present a new way to patch binaries when recompiling from source is not an option.
While it cannot yet replace full-source, we show that decompilation generates recompilable code for most functions.
By focusing on only the vulnerable functions, state-of-the-art decompilers can produce recompilable code that is amenable to source-level code repair tools.
We CGFL to identify a buggy function set, partial decompilation to lift part of the binary to source, where repairs are developed and applied, then generates a \PRD binary addressing the problem.
Our implementation and datasets are available at {{\ourgit}}.

Today's tools are better at finding vulnerabilities than they are at patching them.
We hope our methods will improve that
capacity by leveraging recent advances in source-level \APR.
Although \APR is an active area of research and used in industry,
that potential has not been equally realized for binary code.  \sysacronym using \PRD helps to address these shortfalls.

%% file: references.tex
\bibliography{unified}